\documentclass{aa}
\usepackage{txfonts}
\usepackage{longtable}
\usepackage{graphicx}
\newcommand{\kmprs}  {\mbox{\rm km\,s$^{-1}$}}

\newcommand{\feh} {\mbox{\rm [Fe/H]}}
\newcommand{\fehI} {\mbox{\rm [Fe/H]$_{\rm I}$}}
\newcommand{\fehII} {\mbox{\rm [Fe/H]$_{\rm II}$}}

\newcommand{\ofe} {\mbox{\rm [O/Fe]}}

\newcommand{\nafe} {\mbox{\rm [Na/Fe]}}
\newcommand{\mgfe} {\mbox{\rm [Mg/Fe]}}
\newcommand{\sife} {\mbox{\rm [Si/Fe]}}
\newcommand{\cafe} {\mbox{\rm [Ca/Fe]}}
\newcommand{\tife} {\mbox{\rm [Ti/Fe]}}
\newcommand{\crfe} {\mbox{\rm [Cr/Fe]}}
\newcommand{\mnfe} {\mbox{\rm [Mn/Fe]}}
\newcommand{\nife} {\mbox{\rm [Ni/Fe]}}
\newcommand{\cufe} {\mbox{\rm [Cu/Fe]}}
\newcommand{\znfe} {\mbox{\rm [Zn/Fe]}}
\newcommand{\yfe} {\mbox{\rm [Y/Fe]}}
\newcommand{\bafe} {\mbox{\rm [Ba/Fe]}}
\newcommand{\xx} {\mbox{\rm [X1/X2]}}
\newcommand{\bay} {\mbox{\rm [Ba/Y]}}
\newcommand{\yba} {\mbox{\rm [Y/Ba]}}
\newcommand{\alphafe} {\mbox{\rm [$\alpha$/Fe]}}

\newcommand{\fracxx} {\mbox{${\rm [\frac{X1}{X2}]}$}}
\newcommand{\teff}  {\mbox{$T_{\rm eff}$}}

\newcommand{\logg}  {\mbox{{\rm log}\,$g$}}

\newcommand{\CI} {\ion{C}{i}}
\newcommand{\OI} {\ion{O}{i}}
\newcommand{\MgI} {\ion{Mg}{i}}
\newcommand{\NaI} {\ion{Na}{i}}
\newcommand{\SiI} {\ion{Si}{i}}

\newcommand{\CaI} {\ion{Ca}{i}}
\newcommand{\TiI} {\ion{Ti}{i}}

\newcommand{\CrI} {\ion{Cr}{i}}

\newcommand{\MnI} {\ion{Mn}{i}}
\newcommand{\FeI} {\ion{Fe}{i}}
\newcommand{\FeII} {\ion{Fe}{ii}}
\newcommand{\NiI} {\ion{Ni}{i}}
\newcommand{\CuI} {\ion{Cu}{i}}
\newcommand{\ZnI} {\ion{Zn}{i}}

\newcommand{\YII} {\ion{Y}{ii}}
\newcommand{\BaII} {\ion{Ba}{ii}}

\newcommand{\Vtotal}   {\mbox{$V_{\rm total}$}}
\newcommand{\Rapo}   {\mbox{$R_{\rm apo}$}}

\newcommand{\Zmax}  {\mbox{$Z_{\rm max}$}}

\def\ltsima{$\; \buildrel < \over \sim \;$}
\def\simlt{\lower.5ex\hbox{\ltsima}}
\def\gtsima{$\; \buildrel > \over \sim \;$}
\def\simgt{\lower.5ex\hbox{\gtsima}}
\begin{document}

\title{Two distinct halo populations in the solar neighborhood. II. 
\thanks{Based on observations made with the Nordic Optical Telescope
on La Palma, and on data from the European Southern Observatory
ESO/ST-ECF Science Archive Facility (programs 65.L-0507,
67.D-0086, 67.D-0439, 68.D-0094, 68.B-0475, 69.D-0679,
70.D-0474, 71.B-0529, 72.B-0585, 76.B-0133 and  77.B-0507).}
\fnmsep\thanks{Tables 1, 2, and part of Table 3 are provided as online material
and are also available at the CDS via anonymous ftp to 
{\tt cdsarc.u-strasbg.fr (130.79.128.5a}) or via
{\tt http://cdsarc.u-strasbg.fr/viz-bin/qcat?/A+A/XXX/xxx}.
The complete Table 3 is available in electronic form
at the CDS.}}

\subtitle{Evidence from stellar abundances of Mn, Cu, Zn, Y, and Ba.}

\author{P.E.~Nissen \inst{1} \and W.J.~Schuster \inst{2}}


\institute{
Department of Physics and Astronomy, University of Aarhus, DK--8000
Aarhus C, Denmark.
\email{pen@phys.au.dk}
\and Observatorio Astron\'{o}mico Nacional, Universidad Nacional Aut\'{o}noma
de M\'{e}xico, Apartado Postal 877, C.P. 22800 Ensenada, B.C., M\'{e}xico.
\email{schuster@astrosen.unam.mx}}

\date{Received 1 February 2011 / Accepted 16 March 2011}

\abstract
{Current models of galaxy formation predict that the Galactic halo
was assembled hierarchically. By measuring
abundance ratios in stars it may be possible to identify substructures
in the halo resulting from this process.}
{A previous study of 94 dwarf stars with $-1.6 <$ \feh $< -0.4$ 
in the solar neighborhood has revealed the existence of
two distinct halo populations with a systematic difference in
[$\alpha$/Fe] at a given metallicity. In continuation of that
work, abundances of Mn, Cu, Zn, Y, and Ba are determined 
for the same sample of stars.} 
{Equivalent widths of atomic lines are measured from high resolution 
VLT/UVES and NOT/FIES spectra and used to derive
abundance ratios from an LTE analysis based on MARCS model atmospheres.
The analysis is made relative to two
thick-disk stars, \object{HD\,22879} and \object{HD\,76932},
such that very precise differential values are obtained.}
{Systematic differences between the `high-$\alpha$' and `low-$\alpha$' halo populations
are found for [Cu/Fe], [Zn/Fe], and [Ba/Y], whereas there is no
significant difference in the case of [Mn/Fe]. At a given metallicity, [Cu/Fe]
shows a large scatter that is closely correlated with a corresponding
scatter in [Na/Fe] and [Ni/Fe].}
{The metallicity trends of [Cu/Fe], [Zn/Fe], and [Ba/Y] can be explained from
existing nucleosynthesis calculations if the
high-$\alpha$ stars formed in regions with such a
high star formation rate that only massive stars and Type II supernovae 
contributed to the chemical enrichment. 
The low-$\alpha$ stars, on the other hand, most likely
originate from systems with a slower chemical evolution,
characterized by additional enrichment from
Type Ia supernovae and low-mass AGB stars.}


\maketitle

\section{Introduction}
\label{sect:introduction} 
According to cosmological $\Lambda$CDM simulations, 
large galaxies such as the Milky Way were formed 
hierarchically. Evidence of spatial and kinematical substructures
in the Galactic halo resulting from this process 
has indeed been found as reviewed by Helmi (\cite{helmi08})
and Klement (\cite{klement10}). 
Elemental abundance ratios of halo stars may also
be used to probe this formation process
by so-called `chemical tagging' of the `building blocks' 
(Freeman \& Bland-Hawthorn \cite{freeman02}). The F, G, and K stars 
are of particular interest in this connection, because 
their atmospheric composition are likely to provide a `fossil'
record of the composition of the gas from which the stars once were formed. 
As found by Korn et al. (\cite{korn07}) from a comparison of dwarf and giant stars
in the globular cluster NGC\,6397, the atmospheric abundances of 
heavy elements like Mg, Ca, Ti, and Fe in old dwarf stars
may be somewhat decreased by diffusion processes, but the
ratios between these elements are practically unchanged.

Most high-precision studies of abundance ratios in stars are 
limited to the solar region of the Galaxy. This means that halo
stars must be identified by their kinematics, e.g. by having space
velocities with respect to the local standard of rest (LSR)
well above the characteristic velocities of thin and thick-disk stars.
Several such studies have focused on possible correlations between
kinematics and \alphafe , where $\alpha$ refers to the
abundance of $\alpha$-capture elements like Mg, Si, Ca and Ti.
Fulbright (\cite{fulbright02}) finds evidence that stars with high
values of the total space velocity relative to the LSR,
$\Vtotal > 300$\,\kmprs , tend to have lower values of \alphafe\ than
stars with $150 < \Vtotal < 300$\,\kmprs . Stephens \& Boesgaard
(\cite{stephens02}) show that \alphafe\ is correlated with the apogalactic
orbital distance (\Rapo ) in the sense that the outermost stars have the lowest
values of \alphafe . Gratton et al. (\cite{gratton03}) 
divided their sample into two populations according to 
kinematics: $i)$ a `dissipative' component, which comprises
thick-disk stars and prograde-rotating halo stars, and $ii)$ an
`accretion' component consisting of retrograde-rotating halo
stars. The accretion component has lower values and a larger
scatter for \alphafe\ than the dissipative component. This
has been confirmed by Jonsell et al. (\cite{jonsell05}). Furthermore,
a recent study of Ishigaki et al. (\cite{ishigaki10}) shows that
in the metallicity range $-2 < \feh < -1$, the \mgfe\ ratio for stars
reaching a maximum vertical distance  $\Zmax > 5$\,kpc
above or below the Galactic plane in their orbits tend to be about
0.1\,dex lower than \mgfe\ for stars with $\Zmax < 5$\,kpc. 

The $\alpha$-elements are mainly produced
during Type II supernovae (SNe\,II) explosions of massive stars 
on a relatively short time scale, $\sim \! 10^7$ years, whereas iron 
is also produced by Type Ia SNe (SNe\,Ia) on a longer time scale, 
i.e. from about $10^8$ to more than $10^{9}$ years (Maoz et al. \cite{maoz10}).
The differences in \alphafe\ may therefore be explained in
terms of differences in the star formation rate (SFR). The outer halo
stars may originate from regions characterized
by a relatively slow SFR with both SNe\,Ia and SNe\,II
contributing to the chemical evolution, whereas the inner stars
come from regions with such a fast chemical evolution that only
SNe\,II have contributed. Differences in the initial mass 
function may, however, also play a role, because low-mass SNeII produce
lower values of \alphafe\ than high-mass SNeII
(e.g. Kobayashi et al. \cite{kobayashi06}).

It is unclear from the works cited above
if there is a dichotomy in the distribution of \alphafe\ 
or a more continuous change in \alphafe\ as a function of \Rapo\
or \Zmax . 
In a recent study, the authors of the present paper
(Nissen \& Schuster \cite{nissen10}, hereafter NS10)
have, however, found evidence for the existence of two distinct
halo populations in the solar neighborhood.
Abundances were determined for a sample of 94 
dwarf and subgiant stars with halo or thick-disk kinematics  
selected from Str\"{o}mgren photometry (Schuster et al.
\cite{schuster06}) to have metallicities
$-1.6 < \feh < -0.4$ and to lie in a fairly narrow temperature range,
$5200 < \teff < 6300$\,K. A differential abundance analysis
of high-resolution spectra enabled us to
determine \alphafe \footnote{$\alpha$ refers to the average
abundance of Mg, Si, Ca, and Ti} with a precision of about 0.02\,dex.
As seen from Fig. 1 of NS10, the halo stars are distributed into two groups
clearly separated in \alphafe\ for the metallicity range
$-1.4 < \feh < -0.7$, i.e. `high-$\alpha$' stars having
\alphafe\ close to 0.30~dex, similar to thick-disk stars, and
`low-$\alpha$' stars with \alphafe\ decreasing from 0.25\,dex at $\feh = -1.4$
to 0.10\,dex at $\feh = -0.7$. Differences in \nafe\ and \nife\
are also present with a remarkably tight correlation between these two ratios.
\crfe , on the other hand, is the same for the two populations.

The halo stars in the NS10 sample were selected to have 
$\Vtotal > 180$\,\kmprs . Yet, the high- and low-$\alpha$ populations
have different kinematical properties as seen from Fig. 3 in NS10.
The high-$\alpha$ stars tend to move on prograde Galactic orbits, whereas 
two-thirds of the low-$\alpha$ stars have retrograde orbits 
with an average Galactic rotational velocity component,
$V_{\rm LSR} \simeq -260$\,\kmprs, close to that of the
$\omega$\,Cen globular cluster (Dinescu et al. \cite{dinescu99}).
Furthermore, the low-$\alpha$ stars have a very 
wide distribution of the radial velocity component, $U$,
but comparatively small values of the $W$ component.
All together, this suggests that the low-$\alpha$ stars have been accreted from
satellite galaxies including the progenitor
galaxy of $\omega$\,Cen (Bekki \& Freeman \cite{bekki03}).
The high-$\alpha$ stars, on the other hand, are more likely 
to be formed `in situ' in the Galactic bulge or disk,
and then `heated' to halo kinematics by the merging satellites, 
as suggested from numerical simulations 
(Zolotov et al. \cite{zolotov09}, \cite{zolotov10};
Purcell et al. \cite{purcell10}; Qu et al. \cite{qu11}).

In order to obtain new insight into this scenario and the possible
connection between $\omega$\,Cen and some of the low-$\alpha$ stars,
our abundance analysis has been extended to include Mn, Cu, Zn,
Y, and Ba.  In the following Sect. \ref{sect:abundances},
we describe how the abundances of these elements are determined 
and evaluate statistical and systematic errors. 
In Sect. \ref{sect:discussion}, it is discussed if the metallicity
trends and differences
in abundance ratios between the two halo populations can be explained
from current knowledge of the nucleosynthesis of elements in stars and supernovae.
Furthermore, the abundance ratios of the low-$\alpha$ stars are
compared with the corresponding ratios in present-day
Milky Way satellite galaxies and the $\omega$\,Cen globular cluster.
Finally, some conclusions are given in Sect. \ref{sect:conclusions}. 

\section{Derivation of abundances} 
\label{sect:abundances}

\subsection{Methods}
\label{sect:methods}
\begin{figure}
\resizebox{\hsize}{!}{\includegraphics{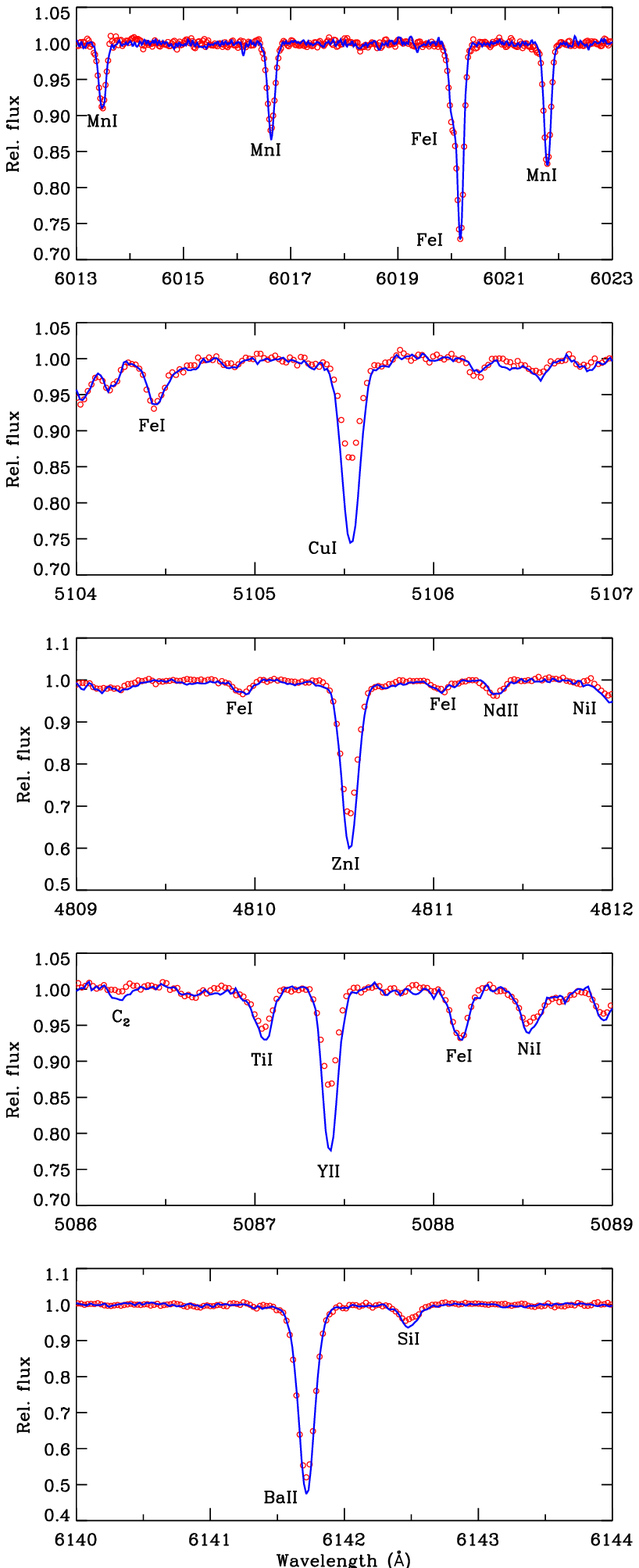}}
\caption{Comparison of the spectra  of a high- and a low-$\alpha$
star with similar atmospheric parameters and \feh . 
The spectrum of 
\object{G\,159-50} (\teff \,=\,5624\,K, \logg \,=\,4.37, \feh
\,=\,$-0.93$, \alphafe \,=\,0.31) is shown with a full drawn
(blue) line, and that of
\object{CD\,$-45\,3283$} (\teff \,=\,5597\,K, \logg \,=\,4.55, \feh
\,=\,$-0.91$, \alphafe \,=\,0.12) with open (red) circles.
The strengths of the Mn and Fe lines are about the same 
in the two stars, whereas the Cu, Zn, Y, and Ba lines are weaker
in the low-$\alpha$ star.}
\label{fig:all-lines}
\end{figure}

Elemental abundances are derived from a model-atmosphere
analysis of equivalent widths (EWs) measured 
in high signal-to-noise (S/N), high-resolution spectra 
observed either with the ESO/VLT UVES
spectrograph or the FIES spectrograph at the Nordic Optical 
Telescope (NOT). Details on wavelength range, resolution, and 
S/N are given in NS10. Figure \ref{fig:all-lines} shows
a comparison of the spectra of a high- and a low-$\alpha$ star
in spectral bands containing some of the Mn, Cu, Zn, Y, and
Ba lines applied in this paper.

As described in NS10, local thermodynamic equilibrium (LTE) is
assumed and 
abundance ratios are determined relative to two bright thick-disk stars,
\object{HD\,22879} and \object{HD\,76932}. They are nearby stars
for which \teff\ can be determined from
photometric colors, and \logg\ via Hipparcos parallaxes.
Using a subset of the spectral lines for which the equivalent widths
could be measured reliably in the solar flux spectrum
(Kurucz et al. \cite{kurucz84}), an abundance analysis
relative to the Sun is carried out. Adopting these 
abundances, an `inverted' analysis leads to the determination of
$gf$-values for all lines. The derived $gf$-values for
\object{HD\,22879} and \object{HD\,76932} agree within $\pm
0.03$\,dex for the large majority of lines; for each line,
the mean value of $gf$ is therefore adopted and used for the
analysis of all program stars. 

This procedure ensures a high internal precision of the abundance ratios.
Relative to the
Sun, the abundances are, however, more uncertain; line blending
and continuum setting are often problematic in the solar spectrum,
and non-LTE effects may be different for the Sun and
a typical program star with $\feh \sim -1$. Thus, 
we find the average difference in Fe abundance derived from 
\FeII\ and \FeI\ lines to be  $\fehII - \fehI = 0.075$\,dex for the  
two standard stars. This difference is 
probably caused by an over-ionization of \FeI\ with respect to LTE in
the metal-poor stars (Asplund \cite{asplund05}). 
\fehII\ is adopted as the metallicity parameter, but
abundance ratios are based on lines from the same ionization
stage, e.g. \mnfe\ is derived from \MnI\ and \FeI\ lines. Hence, if
the non-LTE effects are the same for the two sets of lines,   
\mnfe\ is on the correct scale, but as discussed 
in Sect. \ref{sect:cr-mn} this may not the case. 
Metallicity dependent departures from LTE could 
also affect the derived trends of abundance
ratios as a function of \feh . However, when comparing stars with the
same metallicity and within narrow ranges of \teff\ and \logg ,
non-LTE effects are expected to be the same; the LTE
approximation is therefore adequate in providing reliable differential
abundances at a given metallicity. 
 
For many program stars, colors are affected by
interstellar reddening and the parallaxes are not accurate
enough to determine reliable surface gravities.
Thus, \teff\ is determined from the excitation
balance of weak ($EW < 50$\,m\AA ) \FeI\ lines,
and \logg\ by requiring that the Fe abundances derived from
\FeII\ and \FeI\ lines have the same difference as in
the case of the standard stars, i.e. 0.075 dex.

Model atmospheres are interpolated from the new MARCS grid
(Gustafsson et al. \cite{gustafsson08}), which includes two
sets of models with different values of \alphafe . 
Iteration is performed
until the derived \teff , \logg , \feh , and \alphafe\ are consistent
with the corresponding values of the model atmosphere.

In the following, the determination of the abundances of
Mn, Cu, Zn, Y, and Ba is described in some detail.
Line broadenings caused
by microturbulence and collisional damping are included.
The microturbulence velocity is obtained by minimizing
the dependence of \feh\ on the equivalent widths of \FeI\ lines
with $\chi_{\rm exc} \! > \! 3.0$\,eV. The cross sections for
collisional broadening of \MnI , \CuI , \ZnI , and \BaII\ lines
are taken from Anstee \& O'Mara (\cite{anstee95}),
Barklem \& O'Mara (\cite{barklem97}), and 
Barklem et al. (\cite{barklem00}).
For \YII\ lines, the Uns\"{o}ld (\cite{unsold55})
approximation to the Van der Waals interaction constant, 
enhanced by a factor of two, is adopted.  The
\YII\ lines employed are quite weak ($EW \simlt 60$\,m\AA ),
and are therefore not much affected by possible
errors in the collisional cross sections. 

Abundances of Zn, Y, and Ba are determined from equivalent
widths using the Uppsala EQWIDTH code. For
the Mn and Cu lines, which are significantly affected
by hyperfine splitting (HFS), the BSYN code
is applied to calculate equivalent widths as a function
of abundance. Interpolation to the measured equivalent width
then provides the stellar abundance value.
Results based on UVES and FIES spectra are given in 
Tables \ref{table:UVES} and \ref{table:FIES},
respectively, together with data for \teff , \logg , \feh ,
and \alphafe\ as determined in NS10.
In addition, spectral line data and equivalent widths measured from
UVES spectra of some selected stars (including the prototypes of
a low- and a high-$\alpha$ star, \object{CD\,$-45\,3283$} and
\object{G\,159-50}) are given in Table \ref{table:lines}.
The complete Table \ref{table:lines} with equivalent widths for all
UVES and FIES stars is available at the CDS.

\onltab{1}{
\begin{table*}
\caption[ ]{Atmospheric parameters and abundance ratios
for stars with VLT/UVES spectra.}
\label{table:UVES}
\setlength{\tabcolsep}{0.2cm}
\begin{tabular}{lccccrrrrrr}
\noalign{\smallskip}
\hline\hline
\noalign{\smallskip}
\noalign{\smallskip}
  ID & \teff\ (K)  & \logg  & \feh  & \alphafe  & \mnfe  & \cufe  & \znfe  & 
  \yfe  & \bafe & Class$^{\rm a}$ \\ 
\noalign{\smallskip}
\hline
\noalign{\smallskip}
BD$-$21 3420 &  5808 &     4.26 &    $-$1.13 &     0.31 &   $-$0.32 &   $-$0.31 &    0.17 &    0.11 &    0.02 &  TD       \\
CD$-$33 3337 &  5979 &     3.86 &    $-$1.36 &     0.30 &   $-$0.27 &   $-$0.22 &    0.10 &   $-$0.06 &   $-$0.19 &  TD       \\
CD$-$43 6810 &  5945 &     4.26 &    $-$0.43 &     0.23 &   $-$0.19 &   $-$0.01 &    0.18 &    0.23 &   $-$0.08 &  high-$\alpha$    \\
CD$-$45 3283 &  5597 &     4.55 &    $-$0.91 &     0.12 &   $-$0.31 &   $-$0.52 &   $-$0.03 &   $-$0.20 &   $-$0.14 &   low-$\alpha$    \\
CD$-$51 4628 &  6153 &     4.31 &    $-$1.30 &     0.22 &   $-$0.36 &           &    0.00 &   $-$0.11 &   $-$0.16 &   low-$\alpha$    \\
CD$-$57 1633 &  5873 &     4.28 &    $-$0.90 &     0.07 &   $-$0.34 &   $-$0.69 &   $-$0.10 &   $-$0.27 &   $-$0.14 &   low-$\alpha$    \\
CD$-$61 282  &  5759 &     4.31 &    $-$1.23 &     0.22 &   $-$0.37 &   $-$0.68 &    0.05 &   $-$0.16 &   $-$0.24 &   low-$\alpha$    \\
G05-19     &    5854 &     4.26 &    $-$1.18 &     0.19 &   $-$0.35 &   $-$0.62 &    0.01 &   $-$0.12 &   $-$0.17 &   low-$\alpha$    \\
G05-40     &    5795 &     4.17 &    $-$0.81 &     0.31 &   $-$0.27 &   $-$0.13 &    0.18 &    0.06 &    0.00 &  high-$\alpha$    \\
G18-28     &    5372 &     4.41 &    $-$0.83 &     0.31 &   $-$0.22 &   $-$0.09 &    0.17 &    0.04 &   $-$0.07 &  high-$\alpha$    \\
G18-39     &    6040 &     4.21 &    $-$1.39 &     0.34 &   $-$0.36 &   $-$0.58 &    0.19 &    0.12 &   $-$0.02 &  high-$\alpha$    \\
G20-15     &    6027 &     4.32 &    $-$1.49 &     0.24 &   $-$0.32 &           &    0.11 &   $-$0.14 &   $-$0.22 &  (low-$\alpha$)   \\
G46-31     &    5901 &     4.23 &    $-$0.83 &     0.15 &   $-$0.29 &   $-$0.37 &   $-$0.03 &   $-$0.05 &   $-$0.08 &   low-$\alpha$    \\
G63-26     &    6043 &     4.18 &    $-$1.56 &     0.37 &   $-$0.32 &           &    0.16 &    0.17 &    0.11 & (high-$\alpha$)   \\
G66-22     &    5236 &     4.41 &    $-$0.86 &     0.12 &   $-$0.28 &   $-$0.50 &   $-$0.05 &   $-$0.16 &   $-$0.10 &   low-$\alpha$    \\
G82-05     &    5277 &     4.45 &    $-$0.75 &     0.09 &   $-$0.28 &   $-$0.54 &   $-$0.08 &   $-$0.19 &   $-$0.10 &   low-$\alpha$    \\
G112-43    &    6074 &     4.03 &    $-$1.25 &     0.24 &   $-$0.19 &   $-$0.52 &    0.30 &   $-$0.14 &   $-$0.27 &   low-$\alpha$    \\
G112-44    &    5819 &     4.25 &    $-$1.29 &     0.22 &   $-$0.23 &   $-$0.50 &    0.28 &   $-$0.25 &   $-$0.29 &   low-$\alpha$    \\
G114-42    &    5643 &     4.38 &    $-$1.10 &     0.19 &   $-$0.30 &   $-$0.36 &    0.02 &   $-$0.14 &   $-$0.18 &   low-$\alpha$    \\
G121-12    &    5928 &     4.23 &    $-$0.93 &     0.10 &   $-$0.33 &   $-$0.75 &   $-$0.12 &   $-$0.17 &   $-$0.14 &   low-$\alpha$    \\
G159-50    &    5624 &     4.37 &    $-$0.93 &     0.31 &   $-$0.25 &   $-$0.19 &    0.14 &    0.05 &   $-$0.01 &  high-$\alpha$    \\
G188-22    &    5974 &     4.18 &    $-$1.32 &     0.35 &   $-$0.34 &   $-$0.41 &    0.15 &    0.15 &    0.03 &  high-$\alpha$    \\
HD3567     &    6051 &     4.02 &    $-$1.16 &     0.21 &   $-$0.33 &   $-$0.72 &    0.01 &   $-$0.05 &   $-$0.09 &   low-$\alpha$    \\
HD17820    &    5773 &     4.22 &    $-$0.67 &     0.29 &   $-$0.19 &   $-$0.07 &    0.14 &   $-$0.11 &   $-$0.23 &  TD       \\
HD22879    &    5759 &     4.25 &    $-$0.85 &     0.31 &   $-$0.27 &   $-$0.14 &    0.14 &    0.01 &   $-$0.12 &  TD       \\
HD25704    &    5868 &     4.26 &    $-$0.85 &     0.24 &   $-$0.22 &   $-$0.13 &    0.07 &   $-$0.13 &   $-$0.22 &  TD       \\
HD51754    &    5767 &     4.29 &    $-$0.58 &     0.26 &   $-$0.17 &   $-$0.03 &    0.14 &   $-$0.15 &   $-$0.23 &  high-$\alpha$    \\
HD59392    &    6012 &     3.91 &    $-$1.60 &     0.32 &   $-$0.36 &           &    0.14 &    0.13 &   $-$0.07 &  (low-$\alpha$)   \\
HD76932    &    5877 &     4.13 &    $-$0.87 &     0.29 &   $-$0.25 &   $-$0.18 &    0.10 &    0.03 &   $-$0.07 &  TD       \\
HD97320    &    6008 &     4.19 &    $-$1.17 &     0.28 &   $-$0.24 &   $-$0.24 &    0.09 &   $-$0.09 &   $-$0.19 &  TD       \\
HD103723   &    5938 &     4.19 &    $-$0.80 &     0.14 &   $-$0.32 &   $-$0.50 &   $-$0.07 &   $-$0.16 &   $-$0.11 &   low-$\alpha$    \\
HD105004   &    5754 &     4.30 &    $-$0.82 &     0.14 &   $-$0.21 &   $-$0.22 &    0.04 &   $-$0.15 &   $-$0.16 &   low-$\alpha$    \\
HD106516   &    6196 &     4.42 &    $-$0.68 &     0.29 &   $-$0.23 &   $-$0.13 &    0.18 &   $-$0.05 &   $-$0.18 &  TD       \\
HD111980   &    5778 &     3.96 &    $-$1.08 &     0.34 &   $-$0.34 &   $-$0.32 &    0.15 &    0.16 &    0.07 &  high-$\alpha$    \\
HD113679   &    5672 &     3.99 &    $-$0.65 &     0.32 &   $-$0.23 &   $-$0.07 &    0.17 &   $-$0.01 &   $-$0.17 &  high-$\alpha$    \\
HD114762A  &    5856 &     4.21 &    $-$0.70 &     0.24 &   $-$0.19 &   $-$0.07 &    0.10 &   $-$0.16 &   $-$0.25 &  TD       \\
HD120559   &    5412 &     4.50 &    $-$0.89 &     0.30 &   $-$0.17 &   $-$0.03 &    0.16 &   $-$0.08 &   $-$0.22 &  TD       \\
HD121004   &    5669 &     4.37 &    $-$0.70 &     0.32 &   $-$0.22 &   $-$0.09 &    0.14 &    0.04 &   $-$0.12 &  high-$\alpha$    \\
HD126681   &    5507 &     4.45 &    $-$1.17 &     0.35 &   $-$0.32 &   $-$0.25 &    0.15 &    0.23 &    0.14 &  TD       \\
HD132475   &    5646 &     3.76 &    $-$1.49 &     0.38 &   $-$0.36 &   $-$0.45 &    0.19 &    0.19 &    0.06 & (high-$\alpha$)   \\
HD148816   &    5823 &     4.13 &    $-$0.73 &     0.27 &   $-$0.21 &   $-$0.06 &    0.19 &   $-$0.15 &   $-$0.26 &  high-$\alpha$    \\
HD163810   &    5501 &     4.56 &    $-$1.20 &     0.21 &   $-$0.30 &   $-$0.55 &    0.02 &   $-$0.11 &   $-$0.16 &   low-$\alpha$    \\
HD175179   &    5713 &     4.33 &    $-$0.65 &     0.29 &   $-$0.23 &   $-$0.05 &    0.18 &    0.13 &    0.12 &  TD       \\
HD179626   &    5850 &     4.13 &    $-$1.04 &     0.31 &   $-$0.28 &   $-$0.29 &    0.05 &    0.01 &   $-$0.17 &  high-$\alpha$    \\
HD189558   &    5617 &     3.80 &    $-$1.12 &     0.33 &   $-$0.34 &   $-$0.37 &    0.11 &    0.18 &    0.10 &  TD       \\
HD193901   &    5656 &     4.36 &    $-$1.09 &     0.16 &   $-$0.34 &   $-$0.62 &   $-$0.03 &   $-$0.20 &   $-$0.19 &   low-$\alpha$    \\
HD194598   &    5942 &     4.33 &    $-$1.09 &     0.18 &   $-$0.25 &   $-$0.40 &    0.03 &   $-$0.13 &   $-$0.20 &   low-$\alpha$    \\
HD199289   &    5810 &     4.28 &    $-$1.04 &     0.30 &   $-$0.26 &   $-$0.17 &    0.13 &   $-$0.09 &   $-$0.21 &  TD       \\
HD205650   &    5698 &     4.32 &    $-$1.17 &     0.30 &   $-$0.28 &   $-$0.20 &    0.14 &   $-$0.04 &   $-$0.19 &  TD       \\
HD219617$^{\rm b}$   &    5862 &     4.28 &    $-$1.45 &     0.28 &   $-$0.34 &           &    0.09 &   $-$0.25 &   $-$0.28 &  (low-$\alpha$)   \\
HD222766   &    5334 &     4.27 &    $-$0.67 &     0.30 &   $-$0.24 &   $-$0.06 &    0.21 &   $-$0.01 &   $-$0.09 &  high-$\alpha$    \\
HD241253   &    5831 &     4.31 &    $-$1.10 &     0.29 &   $-$0.31 &   $-$0.26 &    0.14 &    0.07 &   $-$0.04 &  TD       \\
HD284248   &    6135 &     4.25 &    $-$1.57 &     0.27 &   $-$0.33 &           &    0.09 &   $-$0.20 &   $-$0.31 &  (low-$\alpha$)   \\
\noalign{\smallskip}
\hline
\end{tabular}

\begin{list}{}{}
\item[$^{\rm a}$]
Classification as thick disk (TD), low-$\alpha$, or high-$\alpha$.
For halo stars with $\feh < -1.4$, the classification
is uncertain and given in parentheses.
\end{list}

\begin{list}{}{}
\item[$^{\rm b}$]
As noted by Takeda \& Takada-Hidai (\cite{takeda10}), \object{HD\,219617}
is a double star comprising $V = 8.77$ and 9.08 mag. components with a
separation of 0.8\,arcsec and nearly identical spectra. The UVES spectrum,
obtained with a 0.8\,arcsec slit under rather poor seeing conditions, is
an unspecified average of the two spectra.
\end{list}

\end{table*}
}

\onltab{2}{
\begin{table*}
\caption[ ]{Atmospheric parameters and abundance ratios
for stars with NOT/FIES spectra.}
\label{table:FIES}
\setlength{\tabcolsep}{0.2cm}
\begin{tabular}{lccccrrrrrr}
\noalign{\smallskip}
\hline\hline
\noalign{\smallskip}
\noalign{\smallskip}
  ID & \teff\ (K)  & \logg  & \feh  & \alphafe  & \mnfe  & \cufe  & \znfe  &
  \yfe  & \bafe & Class$^{\rm a}$ \\
\noalign{\smallskip}
\hline
\noalign{\smallskip}
G05-36     &    6013 &     4.23 &    $-$1.23 &     0.35 &   $-$0.38 &         &    0.12 &    0.16 &    0.09 &  high-$\alpha$    \\
G13-38     &    5263 &     4.54 &    $-$0.88 &     0.32 &   $-$0.21 &   $-$0.06 &    0.21 &    0.08 &   $-$0.06 &  high-$\alpha$    \\
G15-23     &    5297 &     4.57 &    $-$1.10 &     0.34 &   $-$0.32 &   $-$0.14 &    0.22 &    0.38 &    0.25 &  high-$\alpha$    \\
G16-20     &    5625 &     3.64 &    $-$1.42 &     0.26 &   $-$0.34 &         &    0.09 &   $-$0.13 &   $-$0.22 &  (low-$\alpha$)   \\
G20-15$^{\rm b}$  & 6072 & 4.36 &    $-$1.41 &     0.21 &   $-$0.33 &         &    0.09 &   $-$0.09 &   $-$0.16 &  (low-$\alpha$)   \\
G21-22     &    5901 &     4.24 &    $-$1.09 &     0.09 &   $-$0.39 &         &   $-$0.09 &   $-$0.33 &   $-$0.34 &   low-$\alpha$    \\
G24-13     &    5673 &     4.31 &    $-$0.72 &     0.29 &   $-$0.25 &   $-$0.16 &    0.21 &   $-$0.01 &   $-$0.09 &  high-$\alpha$    \\
G24-25$^{\rm c}$     &    5828 &     3.86 &    $-$1.40 &     0.35 &   $-$0.37 &         &    0.22 &    0.82 &    1.45 & (high-$\alpha$)   \\
G31-55     &    5638 &     4.30 &    $-$1.10 &     0.29 &   $-$0.26 &   $-$0.21 &    0.11 &   $-$0.02 &   $-$0.15 &  high-$\alpha$    \\
G49-19     &    5772 &     4.25 &    $-$0.55 &     0.27 &   $-$0.19 &   $-$0.06 &    0.18 &   $-$0.15 &   $-$0.19 &  high-$\alpha$    \\
G53-41     &    5859 &     4.27 &    $-$1.20 &     0.23 &   $-$0.38 &         &    0.03 &    0.05 &    0.24 &   low-$\alpha$    \\
G56-30     &    5830 &     4.26 &    $-$0.89 &     0.11 &   $-$0.32 &   $-$0.54 &   $-$0.12 &   $-$0.27 &   $-$0.19 &   low-$\alpha$    \\
G56-36     &    5933 &     4.28 &    $-$0.94 &     0.20 &   $-$0.27 &   $-$0.31 &    0.01 &   $-$0.05 &   $-$0.02 &   low-$\alpha$    \\
G57-07     &    5676 &     4.25 &    $-$0.47 &     0.31 &   $-$0.22 &   $-$0.01 &    0.20 &   $-$0.08 &   $-$0.13 &  high-$\alpha$    \\
G74-32     &    5772 &     4.36 &    $-$0.72 &     0.30 &   $-$0.24 &   $-$0.12 &    0.15 &   $-$0.01 &   $-$0.16 &  high-$\alpha$    \\
G75-31     &    6010 &     4.02 &    $-$1.03 &     0.20 &   $-$0.34 &   $-$0.68 &   $-$0.03 &   $-$0.11 &   $-$0.16 &   low-$\alpha$    \\
G81-02     &    5859 &     4.19 &    $-$0.69 &     0.19 &   $-$0.24 &   $-$0.21 &    0.06 &   $-$0.17 &   $-$0.13 &  high-$\alpha$    \\
G85-13     &    5628 &     4.38 &    $-$0.59 &     0.28 &   $-$0.20 &   $-$0.01 &    0.21 &    0.00 &   $-$0.07 &  high-$\alpha$    \\
G87-13     &    6085 &     4.13 &    $-$1.09 &     0.20 &   $-$0.34 &   $-$0.58 &   $-$0.03 &   $-$0.11 &   $-$0.16 &   low-$\alpha$    \\
G94-49     &    5373 &     4.50 &    $-$0.80 &     0.31 &   $-$0.19 &   $-$0.05 &    0.21 &    0.02 &   $-$0.14 &  high-$\alpha$    \\
G96-20$^{\rm c}$     &    6293 &     4.41 &    $-$0.89 &     0.28 &   $-$0.28 &   $-$0.13 &    0.11 &    0.28 &    0.68 &  high-$\alpha$    \\
G98-53     &    5848 &     4.23 &    $-$0.87 &     0.19 &   $-$0.30 &   $-$0.28 &    0.06 &   $-$0.06 &   $-$0.12 &   low-$\alpha$    \\
G99-21     &    5487 &     4.39 &    $-$0.67 &     0.29 &   $-$0.23 &   $-$0.04 &    0.20 &   $-$0.03 &   $-$0.11 &  high-$\alpha$    \\
G119-64    &    6181 &     4.18 &    $-$1.48 &     0.28 &   $-$0.34 &         &    0.02 &   $-$0.08 &   $-$0.23 &  (low-$\alpha$)   \\
G125-13    &    5848 &     4.28 &    $-$1.43 &     0.27 &   $-$0.38 &         &    0.05 &   $-$0.22 &   $-$0.31 & (high-$\alpha$)   \\
G127-26    &    5791 &     4.14 &    $-$0.53 &     0.24 &   $-$0.20 &   $-$0.04 &    0.21 &   $-$0.13 &   $-$0.14 &  high-$\alpha$    \\
G150-40    &    5968 &     4.09 &    $-$0.81 &     0.16 &   $-$0.20 &   $-$0.28 &    0.01 &   $-$0.16 &   $-$0.13 &   low-$\alpha$    \\
G161-73    &    5986 &     4.00 &    $-$1.00 &     0.16 &   $-$0.36 &         &   $-$0.08 &   $-$0.19 &   $-$0.11 &   low-$\alpha$    \\
G170-56    &    5994 &     4.12 &    $-$0.92 &     0.17 &   $-$0.30 &   $-$0.40 &   $-$0.04 &   $-$0.14 &   $-$0.15 &   low-$\alpha$    \\
G172-61    &    5225 &     4.47 &    $-$1.00 &     0.19 &   $-$0.26 &   $-$0.30 &    0.09 &    0.11 &   $-$0.05 &   low-$\alpha$    \\
G176-53    &    5523 &     4.48 &    $-$1.34 &     0.18 &   $-$0.35 &   $-$0.57 &    0.08 &   $-$0.26 &   $-$0.26 &   low-$\alpha$    \\
G180-24    &    6004 &     4.21 &    $-$1.39 &     0.33 &   $-$0.34 &   $-$0.43 &    0.12 &    0.07 &   $-$0.18 &  high-$\alpha$    \\
G187-18    &    5607 &     4.39 &    $-$0.67 &     0.26 &   $-$0.19 &   $-$0.05 &    0.11 &   $-$0.16 &   $-$0.18 &  high-$\alpha$    \\
G192-43    &    6170 &     4.29 &    $-$1.34 &     0.26 &   $-$0.38 &         &    0.08 &   $-$0.09 &   $-$0.17 &   low-$\alpha$    \\
G232-18    &    5559 &     4.48 &    $-$0.93 &     0.32 &   $-$0.30 &   $-$0.19 &    0.15 &    0.08 &   $-$0.03 &  high-$\alpha$    \\
HD148816$^{\rm b}$ & 5840 & 4.14 &   $-$0.70 &     0.26 &   $-$0.21 &   $-$0.08 &    0.12 & $-$0.14 &   $-$0.28 &   high-$\alpha$    \\
HD159482   &    5737 &     4.31 &    $-$0.73 &     0.30 &   $-$0.23 &   $-$0.06 &    0.14 &   $-$0.09 &   $-$0.23 &  high-$\alpha$    \\
HD160693   &    5714 &     4.27 &    $-$0.49 &     0.19 &   $-$0.21 &   $-$0.05 &    0.14 &   $-$0.14 &   $-$0.09 &  high-$\alpha$    \\
HD177095   &    5349 &     4.39 &    $-$0.74 &     0.31 &   $-$0.24 &   $-$0.10 &    0.19 &    0.06 &   $-$0.03 &  high-$\alpha$    \\
HD179626$^{\rm b}$ & 5855 & 4.19 &   $-$1.00 &     0.32 &   $-$0.31 &   $-$0.23 &    0.08 &    0.06 &   $-$0.14 &   high-$\alpha$    \\
HD189558$^{\rm b}$ & 5623 & 3.81 &   $-$1.12 &     0.35 &   $-$0.32 &   $-$0.36 &    0.13 &    0.18 &      0.08 &   TD       \\
HD193901$^{\rm b}$ & 5676 & 4.41 &   $-$1.07 &     0.17 &   $-$0.34 &   $-$0.61 &    0.00 &   $-$0.23 & $-$0.24 &   low-$\alpha$    \\
HD194598$^{\rm b}$ & 5926 & 4.32 &   $-$1.08 &     0.20 &   $-$0.28 &   $-$0.36 &    0.01 &   $-$0.14 & $-$0.20 &   low-$\alpha$    \\
HD230409   &    5318 &     4.54 &    $-$0.85 &     0.27 &   $-$0.16 &   $-$0.02 &    0.16 &   $-$0.08 &   $-$0.17 &  high-$\alpha$    \\
HD233511   &    6006 &     4.23 &    $-$1.55 &     0.34 &   $-$0.44 &         &    0.09 &   $-$0.18 &   $-$0.27 & (high-$\alpha$)   \\
HD237822   &    5603 &     4.33 &    $-$0.45 &     0.29 &   $-$0.16 &    0.01 &    0.22 &   $-$0.12 &   $-$0.21 &  high-$\alpha$    \\
HD250792A  &    5489 &     4.47 &    $-$1.01 &     0.24 &   $-$0.24 &   $-$0.42 &    0.03 &   $-$0.12 &   $-$0.14 &   low-$\alpha$    \\
\noalign{\smallskip}
\hline
\end{tabular}

\begin{list}{}{}
\item[$^{\rm a}$]
Classification as thick disk (TD), low-$\alpha$, or high-$\alpha$.
For halo stars with $\feh < -1.4$, the classification
is uncertain and given in parentheses.
\end{list}

\begin{list}{}{}
\item[$^{\rm b}$]
Data are also available from UVES spectra (see Table \ref{table:UVES}).
\end{list}

\begin{list}{}{}
\item[$^{\rm c}$] $s$-process enhanced star.
\end{list}

\end{table*}
}

\onltab{3}{
\clearpage \onecolumn
\begin{longtable}{llcrrrrrrr}
\caption{\label{table:lines} Line data and equivalent widths measured for
the two standard stars (HD\,22879 and HD\,76932), the solar flux,
and the prototypes of low-, and  high-$\alpha$ stars, 
CD\,$-45\,3283$ and G\,159-50.} \\
\noalign{\smallskip}
\hline\hline
\noalign{\smallskip}
  Element & wavelength  & exc.pot.  & log\,$(gf)^{\rm a}$  & HD\,22879  & HD\,76932 & solar flux &
  CD\,$-45\,3283$ & G\,159-50 & etc. \\
      & [\AA ]   &  [eV]  &   & $EW$\,[m\AA ] & $EW$\,[m\AA ] & $EW$\,[m\AA ] &
       $EW$\,[m\AA ] & $EW$\,[m\AA ] &  \\ 
\noalign{\smallskip}
\hline
\noalign{\smallskip}
\noalign{\smallskip}
\endfirsthead
\caption{continued} \\
\hline\hline
\noalign{\smallskip}
  Element & wavelength  & exc.pot.  & log\,$(gf)^{\rm a}$  & HD\,22879  & HD\,76932 & solar flux &
  CD\,$-45\,3283$ & G\,159-50 & etc. \\
      & [\AA ]   &  [eV]  &   & $EW$\,[m\AA ] & $EW$\,[m\AA ] & $EW$\,[m\AA ] &
       $EW$\,[m\AA ] & $EW$\,[m\AA ] &  \\
\noalign{\smallskip}
\hline
\noalign{\smallskip}
\noalign{\smallskip}
\endhead
\hline
\endfoot
   \NaI &    4982.82 &     2.10 & $   -1.00$ &     25.4 &     22.1 &          &      9.9 &     21.5    &    ...  \\
   \NaI &    5682.64 &     2.10 & $   -0.77$ &     38.2 &     35.3 &          &     19.4 &     34.2    &    ...  \\
   \NaI &    5688.21 &     2.10 & $   -0.48$ &     59.8 &     55.8 &          &     32.5 &     54.1    &    ...  \\
   \NaI &    6154.22 &     2.10 & $   -1.59$ &      7.4 &      7.6 &     37.8 &      3.6 &      7.1    &    ...  \\
   \NaI &    6160.74 &     2.10 & $   -1.30$ &     14.0 &     13.2 &     56.9 &      5.4 &     11.8    &    ...  \\
        &            &          &            &          &          &          &          &             &         \\
   \MgI &    4730.03 &     4.34 & $   -2.33$ &          &          &          &          &             &    ...  \\
   \MgI &    5711.09 &     4.34 & $   -1.74$ &     71.0 &     66.4 &    115.5 &     55.9 &     76.2    &    ...  \\
        &            &          &            &          &          &          &          &             &         \\
   \SiI &    5684.48 &     4.95 & $   -1.68$ &     27.6 &     26.0 &     60.7 &     18.6 &     26.3    &    ...  \\
   \SiI &    5690.42 &     4.93 & $   -1.88$ &     20.4 &     18.2 &     54.6 &     13.6 &     18.5    &    ...  \\
   \SiI &    5701.10 &     4.93 & $   -2.06$ &     14.1 &     13.0 &     40.3 &      9.6 &     13.8    &    ...  \\
   \SiI &    6142.48 &     5.62 & $   -1.52$ &     11.2 &     11.4 &     34.6 &      7.4 &     10.0    &    ...  \\
   \SiI &    6145.01 &     5.61 & $   -1.46$ &     13.6 &     12.5 &     38.6 &      9.9 &     12.5    &    ...  \\
   \SiI &    6155.13 &     5.62 & $   -0.89$ &     39.5 &     36.6 &          &     26.4 &     35.1    &    ...  \\
   \SiI &    6237.32 &     5.61 & $   -1.15$ &     24.3 &     23.8 &          &     14.6 &     21.6    &    ...  \\
   \SiI &    6243.81 &     5.61 & $   -1.33$ &     17.3 &     16.8 &     46.8 &     10.2 &     16.2    &    ...  \\
   \SiI &    6244.47 &     5.61 & $   -1.36$ &     16.7 &     15.5 &     45.3 &      9.3 &     15.3    &    ...  \\
   \SiI &    6414.98 &     5.87 & $   -1.08$ &     18.0 &     15.3 &     48.3 &     11.2 &     17.4    &    ...  \\
        &            &          &            &          &          &          &          &             &         \\
   \CaI &    5260.38 &     2.52 & $   -1.79$ &     10.5 &      9.8 &     32.1 &      9.6 &     11.2    &    ...  \\
   \CaI &    5349.46 &     2.71 & $   -0.45$ &     63.1 &     60.9 &          &     61.9 &             &    ...  \\
   \CaI &    5512.98 &     2.93 & $   -0.52$ &     47.9 &     45.1 &          &     45.2 &     49.5    &    ...  \\
   \CaI &    5581.97 &     2.52 & $   -0.63$ &     61.3 &     59.5 &          &     58.9 &     66.1    &    ...  \\
   \CaI &    5590.12 &     2.52 & $   -0.65$ &     59.7 &     58.7 &          &     59.4 &     64.1    &    ...  \\
   \CaI &    6161.28 &     2.52 & $   -1.33$ &     26.9 &     25.8 &     61.5 &     25.1 &     28.7    &    ...  \\
   \CaI &    6166.43 &     2.52 & $   -1.19$ &     34.3 &     32.1 &     72.1 &     32.0 &     36.3    &    ...  \\
   \CaI &    6169.03 &     2.52 & $   -0.85$ &     55.3 &     51.8 &    102.0 &     55.7 &     58.5    &    ...  \\
   \CaI &    6169.55 &     2.52 & $   -0.58$ &     74.0 &     69.1 &          &     74.4 &             &    ...  \\
        &            &          &            &          &          &          &          &             &         \\
   \TiI &    4820.40 &     1.50 & $   -0.34$ &     17.2 &     15.4 &     45.5 &     14.9 &     18.7    &    ...  \\
   \TiI &    4913.61 &     1.87 & $    0.18$ &     23.4 &     21.1 &     55.3 &     19.9 &     25.4    &    ...  \\
   \TiI &    4999.50 &     0.83 & $    0.27$ &     77.4 &     75.2 &          &     76.3 &     81.0    &    ...  \\
   \TiI &    5016.16 &     0.85 & $   -0.55$ &     36.0 &     32.4 &          &     33.0 &     39.6    &    ...  \\
   \TiI &    5022.86 &     0.83 & $   -0.41$ &     44.0 &     40.0 &          &     42.1 &     46.4    &    ...  \\
   \TiI &    5024.84 &     0.82 & $   -0.55$ &     36.9 &     33.9 &          &     35.0 &     41.2    &    ...  \\
   \TiI &    5039.95 &     0.02 & $   -1.12$ &     48.5 &     45.0 &          &     46.8 &     52.3    &    ...  \\
   \TiI &    5147.47 &     0.00 & $   -1.97$ &     13.7 &     11.3 &     39.0 &     13.4 &     16.6    &    ...  \\
   \TiI &    5192.97 &     0.02 & $   -0.99$ &     56.2 &     51.9 &          &     54.1 &     58.8    &    ...  \\
   \TiI &    5210.38 &     0.05 & $   -0.85$ &     62.3 &     57.4 &          &     61.5 &             &    ...  \\
   \TiI &    5866.41 &     1.07 & $   -0.78$ &     18.9 &     16.0 &     52.3 &     15.7 &     21.5    &    ...  \\
   \TiI &    6258.10 &     1.44 & $   -0.34$ &     22.0 &     19.7 &     50.9 &     20.0 &     24.9    &    ...  \\
   \TiI &    6258.70 &     1.46 & $   -0.19$ &     27.7 &     24.4 &          &     24.4 &     30.6    &    ...  \\
   \TiI &    6261.10 &     1.43 & $   -0.46$ &     18.6 &     15.9 &     49.1 &     16.4 &     21.0    &    ...  \\
        &            &          &            &          &          &          &          &             &         \\
   \CrI &    4801.02 &     3.12 & $   -0.19$ &     14.3 &     12.6 &     52.0 &     15.3 &     14.5    &    ...  \\
   \CrI &    4936.33 &     3.11 & $   -0.32$ &     11.6 &     10.2 &     46.4 &     11.7 &     13.1    &    ...  \\
   \CrI &    5296.69 &     0.98 & $   -1.52$ &     47.0 &     43.8 &          &     53.3 &     51.7    &    ...  \\
   \CrI &    5300.74 &     0.98 & $   -2.22$ &     16.2 &     14.3 &          &     20.2 &     19.2    &    ...  \\
   \CrI &    5329.14 &     2.91 & $   -0.17$ &     22.9 &     20.9 &          &     24.5 &     23.7    &    ...  \\
   \CrI &    5345.80 &     1.00 & $   -1.08$ &     67.3 &     66.6 &          &     73.1 &     74.0    &    ...  \\
   \CrI &    5348.32 &     1.00 & $   -1.42$ &     51.3 &     48.7 &          &     57.3 &     56.0    &    ...  \\
        &            &          &            &          &          &          &          &             &         \\
   \MnI &    4783.4  &     2.30 & $   -0.04$ &     70.0 &     67.7 &          &     79.4 &     77.4    &    ...  \\
   \MnI &    4823.5  &     2.32 & $    0.05$ &     71.7 &     68.5 &          &     77.6 &     78.4    &    ...  \\
   \MnI &    6013.5  &     3.07 & $   -0.42$ &     13.4 &     12.9 &     88.3 &     15.1 &     15.3    &    ...  \\
   \MnI &    6016.7  &     3.07 & $   -0.25$ &     19.6 &     18.0 &     97.9 &     21.0 &     21.2    &    ...  \\
   \MnI &    6021.8  &     3.07 & $   -0.10$ &     24.5 &     23.0 &    102.4 &     27.0 &     27.5    &    ...  \\
        &            &          &            &          &          &          &          &             &         \\
   \FeI &    4788.76 &     3.24 & $   -1.74$ &     28.9 &     26.2 &     71.4 &     30.6 &     29.2    &    ...  \\
   \FeI &    4800.64 &     4.14 & $   -1.10$ &     20.6 &     19.3 &          &     22.3 &     21.4    &    ...  \\
   \FeI &    4924.77 &     2.28 & $   -2.17$ &     54.3 &     51.2 &          &     59.2 &     55.5    &    ...  \\
   \FeI &    4939.68 &     0.86 & $   -3.32$ &     65.6 &     62.6 &          &          &             &    ...  \\
   \FeI &    4950.10 &     3.42 & $   -1.56$ &     30.6 &     28.6 &          &     34.2 &     32.0    &    ...  \\
   \FeI &    4962.57 &     4.18 & $   -1.18$ &     17.1 &     15.4 &     59.7 &     17.6 &     16.4    &    ...  \\
   \FeI &    4973.09 &     3.96 & $   -0.77$ &     44.2 &     41.7 &          &     47.2 &     45.9    &    ...  \\
   \FeI &    5044.21 &     2.85 & $   -2.07$ &     32.5 &     29.8 &          &     36.2 &     33.8    &    ...  \\
   \FeI &    5054.67 &     3.64 & $   -2.04$ &      8.6 &      7.4 &     44.1 &      9.0 &     10.5    &    ...  \\
   \FeI &    5067.14 &     4.22 & $   -0.86$ &     27.7 &     26.1 &          &     31.2 &     28.9    &    ...  \\
   \FeI &    5074.74 &     4.22 & $   -0.13$ &     67.6 &     65.5 &          &     75.0 &     71.7    &    ...  \\
   \FeI &    5090.77 &     4.26 & $   -0.49$ &     45.1 &     42.0 &          &     48.6 &     46.2    &    ...  \\
   \FeI &    5109.65 &     4.30 & $   -0.73$ &     30.5 &     28.5 &     84.7 &     33.3 &     31.1    &    ...  \\
   \FeI &    5126.19 &     4.26 & $   -0.85$ &     27.3 &     23.8 &          &     28.1 &     27.7    &    ...  \\
   \FeI &    5127.36 &     0.91 & $   -3.33$ &     62.6 &     61.7 &          &     67.7 &     66.5    &    ...  \\
   \FeI &    5131.47 &     2.22 & $   -2.46$ &     43.2 &     39.8 &          &     46.1 &     45.3    &    ...  \\
   \FeI &    5141.74 &     2.42 & $   -2.23$ &     45.4 &     41.7 &          &     49.1 &     48.0    &    ...  \\
   \FeI &    5145.09 &     2.20 & $   -3.15$ &     15.5 &     13.2 &          &     18.8 &     18.7    &    ...  \\
   \FeI &    5159.05 &     4.28 & $   -0.81$ &     27.9 &     25.8 &          &     29.3 &     28.6    &    ...  \\
   \FeI &    5187.91 &     4.14 & $   -1.24$ &     16.7 &     15.1 &          &     17.8 &     16.7    &    ...  \\
   \FeI &    5198.71 &     2.22 & $   -2.14$ &     58.6 &     58.0 &          &     66.3 &     61.1    &    ...  \\
   \FeI &    5225.52 &     0.11 & $   -4.75$ &     33.7 &     30.4 &     75.0 &     41.5 &     37.6    &    ...  \\
   \FeI &    5228.37 &     4.22 & $   -1.05$ &     19.9 &     18.7 &          &     20.2 &     21.0    &    ...  \\
   \FeI &    5242.49 &     3.63 & $   -0.99$ &     47.9 &     45.5 &          &     50.3 &     48.7    &    ...  \\
   \FeI &    5243.77 &     4.26 & $   -0.99$ &     20.8 &     19.6 &     65.6 &     22.0 &     21.6    &    ...  \\
   \FeI &    5247.05 &     0.09 & $   -4.91$ &     27.4 &     24.0 &          &     34.1 &     32.2    &    ...  \\
   \FeI &    5250.21 &     0.12 & $   -4.90$ &     26.9 &     23.4 &          &     33.7 &     31.4    &    ...  \\
   \FeI &    5250.64 &     2.20 & $   -2.09$ &     61.9 &     61.7 &          &     68.1 &     67.0    &    ...  \\
   \FeI &    5253.46 &     3.28 & $   -1.65$ &     33.2 &     31.0 &          &     38.5 &     35.4    &    ...  \\
   \FeI &    5263.30 &     3.26 & $   -0.97$ &     73.2 &     70.5 &          &          &             &    ...  \\
   \FeI &    5281.79 &     3.04 & $   -0.94$ &     87.6 &     83.4 &          &    101.5 &             &    ...  \\
   \FeI &    5288.52 &     3.69 & $   -1.59$ &     19.0 &     17.3 &     57.7 &     19.7 &     19.5    &    ...  \\
   \FeI &    5307.36 &     1.61 & $   -2.97$ &     47.6 &     45.7 &          &     51.9 &     49.8    &    ...  \\
   \FeI &    5321.10 &     4.43 & $   -1.26$ &      9.0 &      8.4 &     40.9 &     10.0 &      9.3    &    ...  \\
   \FeI &    5322.04 &     2.28 & $   -2.92$ &     20.4 &     18.1 &     64.6 &     24.2 &     22.4    &    ...  \\
   \FeI &    5339.93 &     3.26 & $   -0.80$ &     86.4 &     79.5 &          &          &             &    ...  \\
   \FeI &    5364.87 &     4.44 & $    0.11$ &     73.4 &     68.2 &          &          &             &    ...  \\
   \FeI &    5365.40 &     3.57 & $   -1.25$ &     39.2 &     36.4 &          &     41.3 &     39.7    &    ...  \\
   \FeI &    5367.47 &     4.41 & $    0.19$ &     79.7 &     75.5 &          &          &             &    ...  \\
   \FeI &    5373.70 &     4.47 & $   -0.77$ &     22.1 &     20.6 &     67.9 &     24.4 &     22.4    &    ...  \\
   \FeI &    5379.57 &     3.69 & $   -1.53$ &     21.0 &     19.3 &     65.2 &     22.4 &     22.0    &    ...  \\
   \FeI &    5389.48 &     4.41 & $   -0.45$ &     39.9 &     37.4 &          &     43.6 &     40.6    &    ...  \\
   \FeI &    5393.17 &     3.24 & $   -0.80$ &     87.5 &     79.9 &          &    104.5 &             &    ...  \\
   \FeI &    5398.28 &     4.44 & $   -0.60$ &     30.9 &     28.5 &     78.1 &     32.6 &     30.5    &    ...  \\
   \FeI &    5409.13 &     4.37 & $   -1.06$ &     15.1 &     14.5 &     54.9 &     16.7 &     16.7    &    ...  \\
   \FeI &    5415.20 &     4.39 & $    0.42$ &     97.5 &     92.2 &          &          &             &    ...  \\
   \FeI &    5432.95 &     4.44 & $   -0.63$ &     27.9 &     28.9 &          &     27.8 &     26.8    &    ...  \\
   \FeI &    5445.04 &     4.39 & $    0.00$ &     67.1 &     64.7 &          &     74.8 &     70.2    &    ...  \\
   \FeI &    5473.90 &     4.15 & $   -0.78$ &     35.3 &     33.1 &          &     38.5 &     36.1    &    ...  \\
   \FeI &    5501.47 &     0.96 & $   -3.06$ &     74.8 &     73.5 &          &     80.7 &     78.1    &    ...  \\
   \FeI &    5543.93 &     4.22 & $   -1.03$ &     21.1 &     19.6 &     67.5 &     22.6 &     22.0    &    ...  \\
   \FeI &    5554.89 &     4.55 & $   -0.36$ &     38.1 &     35.8 &          &     41.3 &     39.3    &    ...  \\
   \FeI &    5560.21 &     4.43 & $   -1.04$ &     14.4 &     13.3 &     54.5 &     15.4 &     14.8    &    ...  \\
   \FeI &    5567.39 &     2.61 & $   -2.66$ &     18.1 &     17.1 &          &     21.7 &     20.6    &    ...  \\
   \FeI &    5569.62 &     3.42 & $   -0.63$ &     89.0 &     82.6 &          &          &             &    ...  \\
   \FeI &    5576.09 &     3.43 & $   -0.94$ &     67.2 &     63.6 &          &     75.0 &             &    ...  \\
   \FeI &    5618.63 &     4.21 & $   -1.27$ &     13.6 &     12.9 &     50.3 &     14.3 &     13.9    &    ...  \\
   \FeI &    5624.55 &     3.42 & $   -0.82$ &     76.2 &     70.6 &          &          &             &    ...  \\
   \FeI &    5633.94 &     4.99 & $   -0.23$ &     24.0 &     22.0 &          &     25.3 &     23.6    &    ...  \\
   \FeI &    5638.26 &     4.22 & $   -0.78$ &     32.1 &     30.3 &     83.7 &     35.0 &     33.9    &    ...  \\
   \FeI &    5679.02 &     4.65 & $   -0.71$ &     17.6 &     17.7 &     63.1 &     19.4 &     18.4    &    ...  \\
   \FeI &    5686.53 &     4.55 & $   -0.57$ &     28.2 &     26.6 &          &     31.2 &     28.4    &    ...  \\
   \FeI &    5701.55 &     2.56 & $   -2.21$ &     40.3 &     38.3 &     86.2 &     45.7 &     43.4    &    ...  \\
   \FeI &    5717.83 &     4.28 & $   -1.01$ &     19.5 &     19.6 &          &     22.5 &     20.4    &    ...  \\
   \FeI &    5731.76 &     4.26 & $   -1.11$ &     17.3 &     16.0 &     60.7 &     20.6 &     18.0    &    ...  \\
   \FeI &    5852.22 &     4.55 & $   -1.15$ &      9.5 &      8.7 &     44.1 &      9.5 &      9.1    &    ...  \\
   \FeI &    5856.09 &     4.29 & $   -1.52$ &      6.9 &      6.8 &     33.2 &      7.6 &      7.0    &    ...  \\
   \FeI &    5859.59 &     4.55 & $   -0.58$ &     27.7 &     26.5 &          &     30.3 &     28.1    &    ...  \\
   \FeI &    6003.01 &     3.88 & $   -1.10$ &     33.5 &     31.9 &          &     38.2 &     35.5    &    ...  \\
   \FeI &    6024.06 &     4.55 & $   -0.02$ &     59.0 &     53.8 &          &     68.0 &     59.1    &    ...  \\
   \FeI &    6027.05 &     4.07 & $   -1.13$ &     24.1 &     21.9 &     67.0 &     28.0 &     24.9    &    ...  \\
   \FeI &    6079.01 &     4.65 & $   -0.99$ &     10.8 &     10.1 &     49.0 &     12.6 &     10.1    &    ...  \\
   \FeI &    6082.71 &     2.22 & $   -3.58$ &      7.0 &      5.7 &     35.0 &      7.6 &      8.3    &    ...  \\
   \FeI &    6151.61 &     2.18 & $   -3.31$ &     12.7 &     11.4 &     49.9 &     15.2 &     15.7    &    ...  \\
   \FeI &    6157.72 &     4.07 & $   -1.20$ &     21.1 &     19.8 &     61.9 &     23.5 &     21.3    &    ...  \\
   \FeI &    6165.35 &     4.14 & $   -1.49$ &     10.6 &     10.1 &     44.3 &          &      9.9    &    ...  \\
   \FeI &    6173.33 &     2.22 & $   -2.88$ &     25.5 &     24.3 &     71.2 &     30.5 &     29.5    &    ...  \\
   \FeI &    6180.20 &     2.73 & $   -2.70$ &     14.3 &     13.4 &     55.3 &     18.6 &     17.5    &    ...  \\
   \FeI &    6200.31 &     2.61 & $   -2.43$ &     28.8 &     26.6 &     76.4 &     34.3 &     31.6    &    ...  \\
   \FeI &    6213.43 &     2.22 & $   -2.55$ &     40.9 &     39.1 &     86.2 &     46.1 &     45.2    &    ...  \\
   \FeI &    6219.28 &     2.20 & $   -2.43$ &     48.7 &     45.8 &     96.7 &     54.0 &     51.6    &    ...  \\
   \FeI &    6246.32 &     3.60 & $   -0.87$ &     63.9 &     58.9 &          &     74.4 &     65.7    &    ...  \\
   \FeI &    6252.55 &     2.40 & $   -1.72$ &     75.5 &     73.2 &    128.5 &     82.1 &     82.2    &    ...  \\
   \FeI &    6265.13 &     2.18 & $   -2.52$ &     45.3 &     42.5 &     90.9 &     49.9 &     47.5    &    ...  \\
   \FeI &    6335.33 &     2.20 & $   -2.26$ &     57.3 &     55.2 &          &     65.2 &     62.6    &    ...  \\
   \FeI &    6336.82 &     3.69 & $   -0.93$ &     54.6 &     51.9 &          &     64.5 &     56.7    &    ...  \\
   \FeI &    6355.03 &     2.84 & $   -2.31$ &     24.5 &     22.6 &          &     29.4 &     26.1    &    ...  \\
   \FeI &    6358.68 &     0.86 & $   -4.24$ &     26.3 &     22.9 &     83.9 &     32.8 &     30.3    &    ...  \\
   \FeI &    6393.60 &     2.43 & $   -1.60$ &     80.9 &     78.2 &          &     91.2 &     88.7    &    ...  \\
   \FeI &    6411.65 &     3.65 & $   -0.69$ &     73.1 &     68.1 &          &          &             &    ...  \\
   \FeI &    6421.35 &     2.28 & $   -2.02$ &     67.3 &     63.8 &          &     75.2 &     70.2    &    ...  \\
   \FeI &    6430.85 &     2.18 & $   -2.01$ &     71.3 &     69.7 &          &     78.6 &     75.6    &    ...  \\
        &            &          &            &          &          &          &          &             &         \\
   \FeII &    4993.34 &     2.81 & $   -3.73$ &     15.2 &     15.9 &     41.4 &     11.2 &     12.4    &    ...  \\
   \FeII &    5132.66 &     2.81 & $   -4.17$ &      6.4 &      6.7 &     25.6 &      4.4 &      6.1    &    ...  \\
   \FeII &    5197.57 &     3.23 & $   -2.33$ &     56.0 &     60.8 &          &     47.0 &     48.8    &    ...  \\
   \FeII &    5234.62 &     3.22 & $   -2.29$ &     58.2 &     63.6 &     89.1 &     49.8 &     51.7    &    ...  \\
   \FeII &    5264.80 &     3.23 & $   -3.15$ &     20.8 &     23.2 &     45.8 &     14.7 &     15.8    &    ...  \\
   \FeII &    5284.10 &     2.89 & $   -3.20$ &     32.4 &     34.5 &          &     23.4 &     26.2    &    ...  \\
   \FeII &    5325.55 &     3.22 & $   -3.25$ &     18.2 &     19.6 &     41.4 &     13.0 &     13.4    &    ...  \\
   \FeII &    5414.07 &     3.22 & $   -3.63$ &      8.8 &      9.5 &     27.8 &      6.0 &      6.0    &    ...  \\
   \FeII &    5425.25 &     3.20 & $   -3.31$ &     17.1 &     18.1 &     44.0 &     11.5 &     12.5    &    ...  \\
   \FeII &    5534.84 &     3.24 & $   -2.88$ &     32.7 &     35.4 &     59.1 &     25.3 &     26.6    &    ...  \\
   \FeII &    6084.09 &     3.20 & $   -3.83$ &      6.2 &      7.0 &     20.7 &      3.7 &      3.8    &    ...  \\
   \FeII &    6149.24 &     3.89 & $   -2.75$ &     14.7 &     15.8 &     36.4 &     10.1 &      9.7    &    ...  \\
   \FeII &    6238.38 &     3.89 & $   -2.63$ &     18.2 &     19.8 &     43.8 &     12.2 &     12.5    &    ...  \\
   \FeII &    6247.55 &     3.89 & $   -2.38$ &     27.1 &     29.6 &     52.5 &     19.8 &     21.2    &    ...  \\
   \FeII &    6416.92 &     3.89 & $   -2.75$ &     15.1 &     16.2 &     40.4 &     10.2 &     11.9    &    ...  \\
   \FeII &    6432.67 &     2.89 & $   -3.65$ &     16.4 &     18.1 &     41.0 &     11.5 &     13.1    &    ...  \\
        &            &          &            &          &          &          &          &             &         \\
   \NiI &    4806.98 &     3.68 & $   -0.60$ &     21.5 &     20.2 &          &     17.4 &     20.6    &    ...  \\
   \NiI &    4829.02 &     3.54 & $   -0.33$ &     40.1 &     38.0 &          &     35.3 &     40.3    &    ...  \\
   \NiI &    4831.17 &     3.61 & $   -0.38$ &     34.1 &     31.9 &     77.6 &     28.3 &     34.1    &    ...  \\
   \NiI &    4904.41 &     3.54 & $   -0.19$ &     47.8 &     45.2 &          &     43.3 &     47.8    &    ...  \\
   \NiI &    4913.97 &     3.74 & $   -0.62$ &     18.4 &     18.1 &     60.0 &     15.0 &     19.1    &    ...  \\
   \NiI &    4935.82 &     3.94 & $   -0.31$ &     23.3 &     21.7 &     64.5 &     19.2 &     23.3    &    ...  \\
   \NiI &    4937.34 &     3.61 & $   -0.38$ &     35.0 &     32.4 &          &     29.4 &     34.7    &    ...  \\
   \NiI &    4953.20 &     3.74 & $   -0.61$ &     19.7 &     17.6 &     60.0 &          &     19.2    &    ...  \\
   \NiI &    4998.22 &     3.61 & $   -0.76$ &     18.2 &     17.0 &          &     14.3 &     18.1    &    ...  \\
   \NiI &    5010.93 &     3.63 & $   -0.85$ &     15.2 &     13.9 &     51.7 &     10.8 &     14.4    &    ...  \\
   \NiI &    5035.36 &     3.63 & $    0.12$ &     59.3 &     60.0 &          &     56.0 &     60.7    &    ...  \\
   \NiI &    5081.11 &     3.85 & $    0.22$ &     54.0 &     52.5 &          &     50.2 &     55.3    &    ...  \\
   \NiI &    5082.34 &     3.66 & $   -0.55$ &     24.7 &     23.5 &     75.4 &     20.4 &     25.5    &    ...  \\
   \NiI &    5084.09 &     3.68 & $   -0.04$ &     48.8 &     47.7 &          &     43.1 &     49.3    &    ...  \\
   \NiI &    5099.93 &     3.68 & $   -0.18$ &     40.7 &     40.3 &          &     35.7 &     41.0    &    ...  \\
   \NiI &    5102.96 &     1.68 & $   -2.83$ &     12.3 &     11.6 &          &     11.1 &     14.1    &    ...  \\
   \NiI &    5115.39 &     3.83 & $   -0.14$ &     35.9 &     34.3 &          &     31.4 &     35.8    &    ...  \\
   \NiI &    5155.12 &     3.90 & $   -0.63$ &     14.3 &     13.0 &     53.1 &     11.5 &     14.6    &    ...  \\
   \NiI &    5155.76 &     3.90 & $   -0.11$ &     34.2 &     32.8 &          &     30.0 &     34.8    &    ...  \\
   \NiI &    5176.55 &     3.90 & $   -0.46$ &     19.8 &     17.9 &     59.7 &     14.4 &     19.4    &    ...  \\
   \NiI &    5578.72 &     1.68 & $   -2.74$ &     15.9 &     14.4 &          &     14.0 &     17.4    &    ...  \\
   \NiI &    5587.86 &     1.93 & $   -2.41$ &     18.2 &     17.0 &          &     17.8 &     19.8    &    ...  \\
   \NiI &    5682.20 &     4.10 & $   -0.43$ &     15.0 &     13.8 &     53.4 &     11.0 &     14.5    &    ...  \\
   \NiI &    6086.28 &     4.26 & $   -0.49$ &     10.3 &      9.3 &     44.3 &      8.3 &     10.8    &    ...  \\
   \NiI &    6108.12 &     1.68 & $   -2.61$ &     20.8 &     19.8 &     67.5 &     19.5 &     25.0    &    ...  \\
   \NiI &    6175.36 &     4.09 & $   -0.51$ &     13.3 &     12.8 &     49.2 &     10.6 &     13.6    &    ...  \\
   \NiI &    6176.81 &     4.09 & $   -0.31$ &     19.5 &     18.6 &     67.9 &     16.6 &     20.8    &    ...  \\
        &            &          &            &          &          &          &          &             &         \\
   \CuI &    5105.5  &     1.39 & $   -1.58$ &     31.3 &     25.7 &     95.9 &     18.2 &     32.4    &    ...  \\
   \CuI &    5218.2  &     3.82 & $    0.36$ &     13.7 &     12.2 &     56.5 &      6.8 &     12.1    &    ...  \\
   \CuI &    5782.1  &     1.64 & $   -1.78$ &          &          &          &          &             &    ...  \\
        &            &          &            &          &          &          &          &             &         \\
   \ZnI &    4722.15 &     4.03 & $   -0.50$ &          &          &     70.8 &          &             &    ...  \\
   \ZnI &    4810.53 &     4.08 & $   -0.31$ &     48.8 &     48.5 &     82.8 &     37.7 &     45.7    &    ...  \\
        &            &          &            &          &          &          &          &             &         \\
   \YII &    4883.68 &     1.08 & $    0.01$ &     39.0 &     40.3 &     63.9 &     25.0 &     35.5    &    ...  \\
   \YII &    5087.42 &     1.08 & $   -0.28$ &     26.2 &     27.8 &     51.0 &     14.4 &     23.9    &    ...  \\
        &            &          &            &          &          &          &          &             &         \\
   \BaII &    5853.68 &     0.60 & $   -0.94$ &     38.9 &     41.9 &     70.2 &     32.9 &     39.6    &    ...  \\
   \BaII &    6141.72 &     0.70 & $    0.06$ &     84.2 &     87.7 &    132.5 &     81.7 &     87.5    &    ...  \\
\noalign{\smallskip}
\hline
\end{longtable}

\begin{list}{}{}
\item[$^{\rm a}$]
The large majority of $gf$-values are determined from UVES spectra of the two
standard stars, \object{HD\,22879} and \object{HD\,76932}, as 
described in Sect. \ref{sect:methods}. For three
lines that are only available in the FIES spectra (\MgI\ 4730.0, \CuI\ 5782.1,
and \ZnI\ 4722.2), the $gf$-value is determined so that for the
sample of FIES stars, the mean elemental abundance derived from the line 
agrees with the mean abundance derived from the other lines of the
element.
\end{list}

}

\subsection{Manganese}
\label{sect:mn}
The derived abundances are based on \MnI\ lines at
4783.4, 4823.5, 6013.5, 6016.7, and 6021.8\,\AA\ with
HFS data adopted from Prochaska et al. 
(\cite{prochaska00}). Cunha et al. (\cite{cunha10}) used
the same reference when deriving Mn abundances for giant
stars in $\omega$\,Cen, and tested that the more recent
HFS data of Bergemann et al. (\cite{bergemann07}) changed the
equivalent widths of the $\lambda \, 6013.5$ line by less 
than 1\,\%.

\begin{figure}
\resizebox{\hsize}{!}{\includegraphics{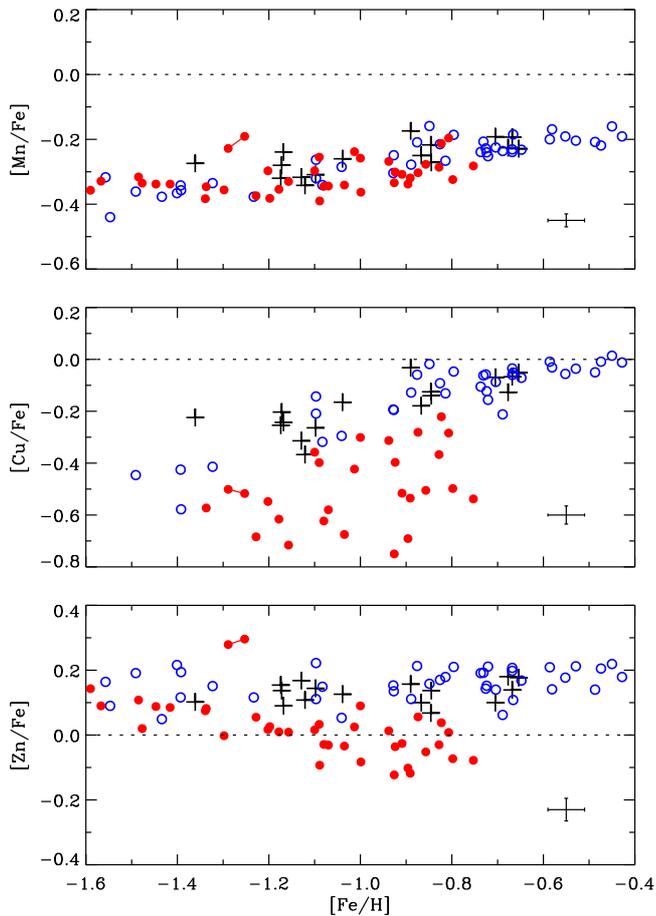}}
\caption{\mnfe , \cufe , and \znfe\ versus \feh .
Stars belonging to the high-$\alpha$ halo population, as
classified from \mgfe , are shown with open circles (blue in the
online version), and the low-$\alpha$ halo population with filled
(red) circles.  The crosses refer to thick-disk stars. Typical  one-sigma error bars
are given in the lower, right corners.  The two components of a 
binary star, \object{G\,112-43} and \object{G\,112-44}, are
connected by a straight line.}
\label{fig:mn.cu.zn-fe}
\end{figure}

The lines near 6000\,\AA\
are quite weak in our stellar spectra ($EW \simlt 50$\,m\AA )
and disappear in the noise at low metallicities ($\feh \simlt -1.4$).
The $\lambda \lambda$\,4783.4,\,4823.5 lines are stronger,
and are disregarded for the cooler and more metal-rich stars
if their equivalent widths exceed 80\,m\AA . Furthermore,
the spectral coverage of the UVES spectra stops at 4780\,\AA ,
so in some cases the $\lambda$\,4783.4 line is lost because of a
high negative radial velocity of the star.  

For a group of
33 intermediate metallicity stars, all five Mn lines 
could be used to determine abundances,
and it was checked that the
Mn abundance derived from a given line is consistent 
with that derived from the other lines. The rms
difference of Mn abundances derived from a single line and
from all lines never exceeded 0.03\,dex. This suggests that 
very precise Mn abundances have been obtained.

The derived \mnfe\ values are shown in Fig. \ref{fig:mn.cu.zn-fe} 
(upper panel)\footnote{For six stars  observed with both UVES
and FIES, only the UVES data are plotted in all figures.
Table \ref{table:UVES-FIES} shows a comparison between the two datasets.}
as a function of \feh\ with different symbols employed for
thick-disk, high-$\alpha$, and low-$\alpha$ halo stars.
The scatter in \mnfe\ at a given \feh\ is small
($\simeq 0.04$\,dex) and the metallicity trends for high- and 
low-$\alpha$ halo stars are nearly the same. The
similarity of \mnfe\ for high- and low-$\alpha$ stars
is also evident from  Fig. \ref{fig:all-lines},
where the three \MnI\ lines near 6000\,\AA\ are compared
for two stars having nearly the same values of \teff ,
\logg , and \feh , but a significant difference in 
\alphafe .

The only two stars, \object{G\,112-43} and \object{G\,112-44},
deviating significantly from
the average trend in Fig. \ref{fig:mn.cu.zn-fe}, are probably 
components in a binary star. 
They are separated by only 11 arcsec on the
sky and have nearly the same radial velocities and proper motions.  
\object{G\,112-43} is at the turn-off, and \object{G\,112-44} 
is a main-sequence star about one magnitude fainter. Both components
belong to the class of low-$\alpha$ halo stars. In all
abundance diagrams in the present paper and in NS10, these two
stars are connected with a straight line.
Their abundances agree very well, but they tend to deviate
from the rest of the low-$\alpha$ stars in several abundance ratios.
Interestingly, the two stars also stand out by having exceptionally
high velocity components perpendicular to the Galactic plane
(see Fig. 8 in NS10).

\subsection{Copper}
\label{sect:cu}
Cu abundances are derived from the equivalent widths
of \CuI\ lines at 5105.5, 5218.2 and 5782.1\,\AA\ with HFS data 
taken from the same source as in the case of Mn 
(Prochaska et al. \cite{prochaska00}). Their data
include isotopic splitting between $^{63}$Cu and $^{65}$Cu  
assuming solar system fractions of 0.69 and 0.31, respectively.
The isotopic splitting is small and the calculated
equivalent width is insensitive to the assumed isotopic 
composition.

The strongest of the three Cu lines, $\lambda \, 5105.5$, is
somewhat affected by blends of C$_2$ molecular lines in the wings,
and care was taken to avoid them by measuring equivalent
widths using the deblending routine in the IRAF {\tt splot} task.
In general, the Cu lines become too faint ($EW \simlt 2$\,m\AA )
to provide reliable abundances in the most metal-poor stars.
Furthermore, the $\lambda \, 5782.1$ line is not available in
the UVES spectra, because it happens to fall in a
gab between the two CCDs in the red arm of the spectrograph.
 
Cu abundances derived from the individual lines agree well.
There is no significant systematic offsets, and a comparison of
Cu abundances derived from the 5105.5 and 5218.2\,\AA\
lines results in a rms deviation of 0.052\,dex.
Cu abundances from the 5218.2 and 5782.1\,\AA\ lines 
in the FIES spectra have a rms deviation of
0.061\,dex. 

As seen from Fig. \ref{fig:mn.cu.zn-fe} (middle panel), 
the high-$\alpha$ and thick-disk stars define a smooth
relation between \cufe\ and \feh .
The low-$\alpha$ stars, on the other hand,
fall below the high-$\alpha$ stars and have a large scatter
in \cufe\ at a given metallicity. As an example of
the difference in the strength of Cu lines 
for high- and low-$\alpha$ stars, Fig. \ref{fig:all-lines}
shows the $\lambda \, 5105.5$ line in the 
spectra of \object{G\,159-50} and \object{CD\,$-45\,3283$}.

\subsection{Zinc}
\label{sect:zn}
Abundances of Zn are determined from the \ZnI\ 
$\lambda \lambda \, 4722.1, \, 4810.5$ lines. The first line
falls outside the range of the UVES spectra, but since
these spectra have higher S/N than those obtained
with FIES, it is estimated that the error of the Zn abundances
is about the same for the two sets of observations. A
comparison of abundances obtained from the two lines
in the FIES spectra results in a rms deviation of 0.045\,dex. 

The relation between \znfe\ and \feh\ is shown in 
Fig. \ref{fig:mn.cu.zn-fe} (lower panel).
There is a clear difference
in the trends of high- and low-$\alpha$ stars. The
thick-disk and high-$\alpha$ halo stars have the same,
near-constant $\znfe \simeq 0.15$ with a rms scatter
of 0.045\,dex around this value. The low-$\alpha$ stars
show a declining trend with increasing \feh , and are
separated from the high-$\alpha$ stars in the
metallicity range $-1.4 < \feh < -0.7$, except for
the components of the binary star.   

\subsection{Yttrium}
\label{sect:y}
The yttrium abundances are based on the \YII\ lines at
4883.7 and 5087.4\,\AA . Being an odd-Z element ($Z = 39$)
HFS is present, but as discussed by Hannaford et al.
(\cite{hannaford82}), the splitting of the hyperfine components
is less than 1\,m\AA\  and can be safely neglected in
connection with the determination of Y abundances.
The equivalent widths of the lines
range from about 4 to 60\,m\AA\ in our sample of stars, and
the Y abundances derived from the two lines agree with 
a rms difference of 0.044\,dex.  From
the outset, \YII\ lines at 5123.2 and 5200.4\,\AA\ were also
included, but it turned out that the corresponding Y abundances   
deviate from the $\lambda \lambda \, 4883.7, \, 5087.4$ abundances
in a systematic way as a function of line strength indicating
the presence of undetected blends.

\begin{figure}
\resizebox{\hsize}{!}{\includegraphics{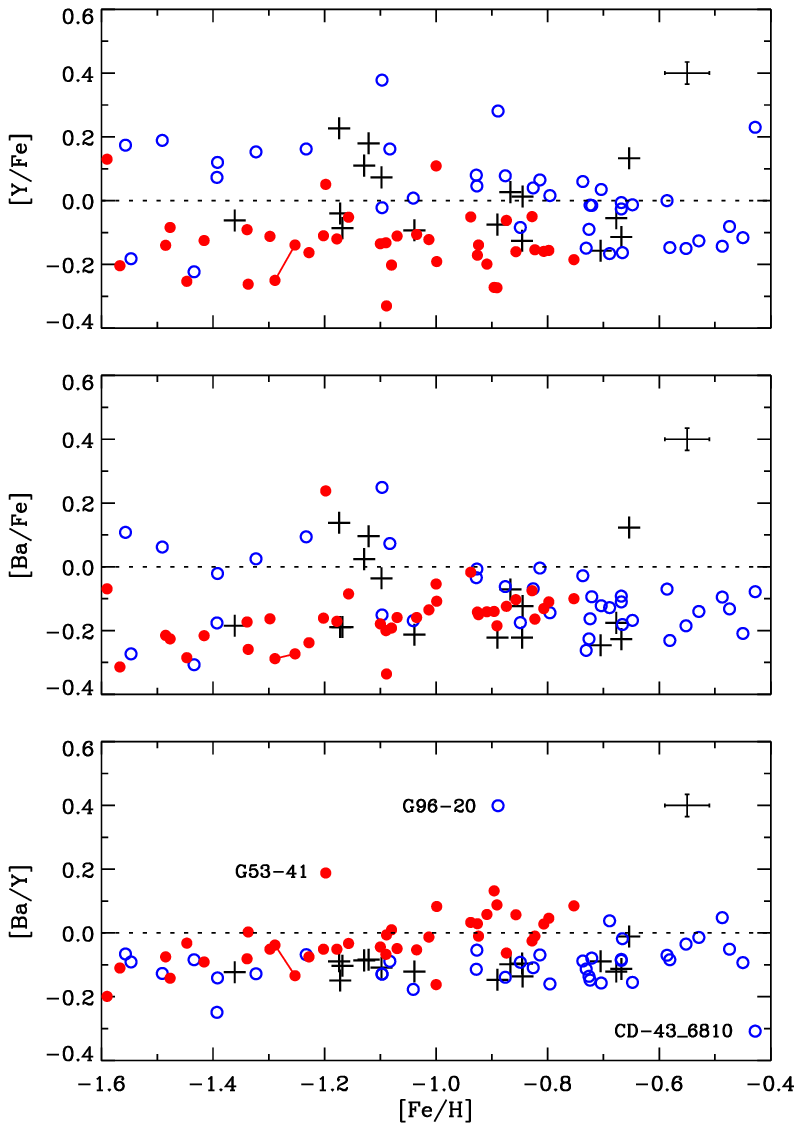}}
\caption{\yfe , \bafe , and \bay\ versus \feh\ with
the same symbols as in Fig. \ref{fig:mn.cu.zn-fe}.}
\label{fig:y.ba.bay-fe}
\end{figure}

Fig. \ref{fig:y.ba.bay-fe} (upper panel) shows \yfe\ as
a function of \feh . Although the low-$\alpha$ stars
tend to have lower \yfe\ values than the high-$\alpha$ stars,
there is considerable scatter in \yfe ; several of
the thick-disk and high-$\alpha$ stars fall among the 
low-alpha stars. In addition, it should be noted that
one star, \object{G\,24-25}, has $\yfe = 0.82$ and falls
above the upper limit of the figure. This star is also
very overabundant in Ba and other $s$-process elements
with an abundance pattern that may be explained by mass
transfer from an AGB companion (L. Shu et al., in preparation).
According to Latham et al. (\cite{latham02}), \object{G\,24-25} 
is an SB1 spectroscopic binary with an orbital period of 
9.45\,years. 

\subsection{Barium}
\label{sect:ba}
Abundances of Ba are determined from the $\lambda \lambda \, 
5853.7, \, 6141.7$ \BaII\ lines. Odd numbered Ba isotopes exhibit
hyperfine splitting, and it was therefore first investigated
if this has any significant effect on the derived abundances.
Adopting HFS data from McWilliam (\cite{mcwilliam98}), profiles and
equivalent widths were calculated for some representative stars.
Contrary to the case of the \BaII\ line at 4554.0\,\AA\
(e.g. Collet et al. \cite{collet09}), the HFS effect on the
$\lambda \lambda \, 5853.7, \, 6141.7$ lines is small and 
changes the derived Ba abundances  
by less than 0.01\,dex even if one assumes a pure $r$-process
distribution of the isotopes instead of a solar system distribution.
Thus, the abundance analysis was carried out neglecting 
hyperfine splitting. 
 
The $\lambda \, 5853.7$ \BaII\ line is suitable for precise abundance 
determination by having equivalent widths from about 10 to 60\,m\AA\
in the majority of our stars. The $\lambda \, 6141.7$ line
has equivalent widths typically in the 
range  50 - 100\,m\AA , and is therefore more sensitive to the 
adopted microturbulence and damping constant. Nevertheless, 
the two sets of Ba abundances show a satisfactory agreement;
the rms difference is 0.058\,dex.

Fig. \ref{fig:y.ba.bay-fe} (middle panel) shows \bafe\
versus \feh . Like in the case of \yfe , there is 
some mixing of high- and low-$\alpha$ stars. 
The $s$-process rich star, \object{G\,24-25}, has $\bafe = 1.45$
and falls far above the upper limit of the figure. The same 
is the case with \object{G\,96-20}, a mildly $s$-process rich star,
which has $\bafe = 0.68$ and $\yfe = 0.28$.

Interestingly, there is a correlation between the scatter in \yfe\ and  
\bafe\ at a given metallicity. This is evident from the lower panel
of Fig. \ref{fig:y.ba.bay-fe}, where \bay\ is plotted  vs.
\feh .  The high- and low-$\alpha$ stars show  well-defined
trends of \bay\ with increasing separation as a function of increasing \feh .
Still, a few stars deviate very significantly from these trends.
\object{G\,24-25} with $\bay = 0.64$ falls above the limit of the figure,
whereas the other $s$-process rich star, \object{G\,96-20}, has
$\bay = 0.40$. Other deviating stars 
are \object{G\,53-41} (Na-rich according to NS10) and
\object{CD\,$-$43\,6810}.

\subsection{Statistical errors}
\label{sect:errors}
An estimate of the statistical errors of the various abundance
ratios can be obtained by comparing values derived from UVES and FIES
spectra for six stars observed with both instruments.
As the atmospheric parameters are determined spectroscopically, 
the comparison will include the effect of
the uncertainties in \teff , \logg , and \feh\ on the abundance ratios.

\begin{table*}
\caption[ ]{Differences (UVES -- FIES) of atmospheric parameters and abundance
ratios determined  from spectra of six stars observed with both instruments.}
\label{table:UVES-FIES}
\setlength{\tabcolsep}{0.20cm}
\begin{tabular}{lrrrrrrrrr}
\noalign{\smallskip}
\noalign{\smallskip}
\hline\hline
\noalign{\smallskip}
    & $\Delta \teff$\,[K] & $\Delta \logg$ & $\Delta \feh$ & $\Delta \mnfe$ & $\Delta \cufe$ & $\Delta \znfe$ & $\Delta \yfe$ & $\Delta \bafe$  \\
\noalign{\smallskip}
\hline
\noalign{\smallskip}
 \object{G\,20-15}  & $-45$ & $-0.04$ & $-0.07$ & $+0.011$ &          & $+0.016$ & $-0.046$ & $-0.060$   \\
\object{HD\,148816} & $-17$ & $-0.01$ & $-0.03$ & $-0.001$ & $+0.021$ & $+0.068$ & $-0.011$ & $+0.016$   \\
\object{HD\,179626} & $ -5$ & $-0.06$ & $-0.04$ & $+0.023$ & $-0.062$ & $-0.026$ & $-0.046$ & $-0.034$   \\
\object{HD\,189558} & $ -6$ & $-0.01$ & $ 0.00$ & $-0.025$ & $-0.008$ & $-0.018$ & $+0.001$ & $+0.014$   \\
\object{HD\,193901} & $-20$ & $-0.05$ & $-0.02$ & $-0.001$ & $-0.018$ & $-0.033$ & $+0.032$ & $+0.044$   \\
\object{HD\,194598} & $+16$ & $+0.01$ & $-0.01$ & $+0.020$ & $-0.035$ & $+0.020$ & $+0.008$ & $+0.003$   \\
\hline
\noalign{\smallskip}
Average($\Delta$)   & $-13$ & $-0.03$ & $-0.03$ & $+0.005$ & $-0.020$ & $+0.004$ & $-0.010$ & $-0.003$   \\
    rms($\Delta$)   &   23  &   0.04  &   0.04  &   0.017  &   0.034  &   0.035  &   0.030  &   0.035     \\
\noalign{\smallskip}
\hline
\end{tabular}
\end{table*}

The results from this comparison are given in Table \ref{table:UVES-FIES}.
The six stars are representative for the whole sample of program
stars. \object{G\,20-15} is one of the most metal poor stars, 
\object{HD\,148816} and \object{HD\,179626} belong to the metal-rich end,
and \object{HD\,189558}, \object{HD\,193901} and \object{HD\,194598} 
have intermediate metallicities. As seen from the table,
the systematic offsets between the UVES and FIES results are small and
can be neglected. The rms scatters of the differences are on the same
order of size as the errors of the average abundances estimated from
comparing abundances derived from individual lines. Thus, the uncertainty
of the atmospheric parameters is not adding much to the error budget, which can
be ascribed to the fact that we are deriving abundance ratios from lines
belonging to the same ionization stage.

On the basis of Table \ref{table:UVES-FIES} and the line-to-line scatter
of derived abundances, we  estimate the one-sigma error of \mnfe\ to be
0.025\,dex and the errors of other ratios to be 0.035\,dex. It is emphasized
that these are statistical errors that can be applied when comparing
relative abundance ratios for stars at a given metallicity.
The absolute values and the trends of
abundance ratios as a function of metallicity may be affected by 
systematic errors caused by non-LTE effects as mentioned in Sect. 2.1.

\section{Discussion}
\label{sect:discussion}
In this section, the trends and correlations in the elemental
abundance ratios for high- and low-$\alpha$ halo stars
are discussed in terms of nucleosynthesis calculations.
Furthermore, the abundance ratios for the low-alpha stars are compared
with ratios determined for dwarf galaxies and the $\omega$\,Cen
globular cluster in an attempt to find evidence for a connection.

\begin{figure}
\resizebox{\hsize}{!}{\includegraphics{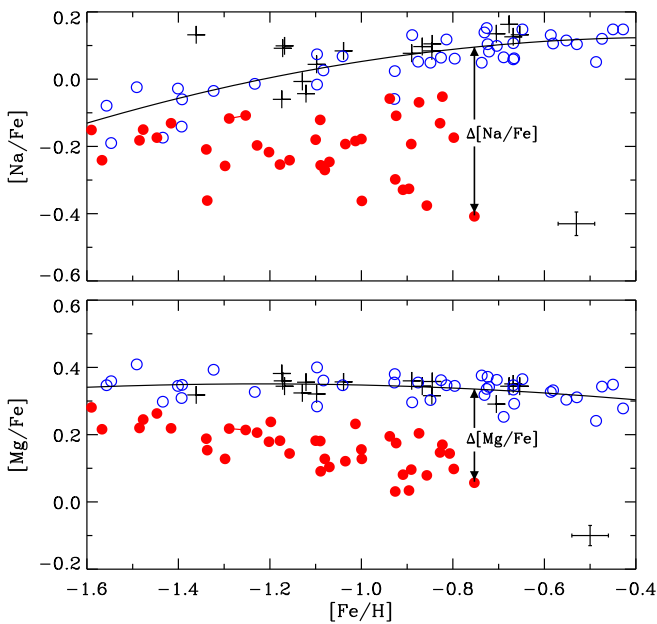}}
\caption{Definition of $\Delta \nafe$ and $\Delta \mgfe$.
The same symbols as in Fig. \ref{fig:mn.cu.zn-fe} are
employed. The lines show quadratic fits to the
distribution of thick-disk and high-$\alpha$ halo stars.}
\label{fig:na.mg-fe}
\end{figure}

\subsection{Differences and correlations between abundance ratios}  
\label{sect:dif-cor}
As seen from Figs. \ref{fig:mn.cu.zn-fe} and \ref{fig:y.ba.bay-fe},
the high- and low-$\alpha$ halo stars begin to separate
in \cufe , \znfe , and \bay\ at $\feh \sim -1.5$
and reach the largest separation for the highest metallicities
of the low-$\alpha$ stars. A similar behavior of
\nafe , \mgfe , \sife , \cafe , \tife , and \nife\ is found in
NS10. Furthermore, the low-$\alpha$ stars tend to have a larger
scatter of these abundance ratios than the high-$\alpha$ stars.
\crfe\ and \mnfe , on the other hand, show little, if any, difference
between the two halo populations, and the scatter in these ratios
is about the same for the two groups.

In order to quantify the differences between the two halo
populations, we have made quadratic fits to the trends of abundance ratios
as a function of \feh\ for the thick-disk and high-$\alpha$ halo stars.
As an example, Fig. \ref{fig:na.mg-fe} shows the fits to \nafe\ and \mgfe .
For a given star, the deviations from the fits, 
$\Delta \nafe$ and $\Delta \mgfe$, are defined as shown on the figure.
Corresponding quadratic fits and deviations are calculated for the other
element ratios. Table \ref{table:ratios}
gives the mean value and the rms scatter of the deviations for the metallicity
range $-1.1 \le \feh < -0.7$, where the largest separation of the
two populations occurs. The high-$\alpha$ group includes 18 halo stars
and 7 thick-disk stars, and the low-$\alpha$ group comprises 21 halo stars,
except in the case of copper, for which two stars have too weak \CuI\ lines to
allow a reliable determination of the Cu abundance.

\begin{table}
\caption[ ]{Mean value and rms scatter of deviations in
abundance ratios for 
high- and low-$\alpha$ stars with $-1.1 \le \feh < -0.7$.} 
\label{table:ratios}
\setlength{\tabcolsep}{0.20cm}
\begin{tabular}{lccccc}
\noalign{\smallskip}
\hline\hline
\noalign{\smallskip}
       & \multicolumn{2}{c}{High-$\alpha$ group} & & \multicolumn{2}{c}{Low-$\alpha$ group} \\
 \xx  & $<\!\Delta \fracxx\!>$  & $\sigma (\Delta \fracxx )$ & & $<\!\Delta \fracxx\!>$ & $\sigma (\Delta \fracxx )$ \\
\noalign{\smallskip}
\hline
\noalign{\smallskip}
 \nafe  & 0.000 & 0.041 &  & $-0.277$ & 0.110 \\
 \mgfe  & 0.002 & 0.029 &  & $-0.217$ & 0.055 \\
 \sife  & 0.001 & 0.028 &  & $-0.175$ & 0.040 \\
 \cafe  & 0.002 & 0.030 &  & $-0.072$ & 0.046 \\
 \tife  & 0.003 & 0.030 &  & $-0.137$ & 0.056 \\
 \crfe  & 0.004 & 0.032 &  & $-0.031$ & 0.029 \\
 \mnfe  & 0.003 & 0.036 &  & $-0.043$ & 0.042 \\
 \nife  & 0.001 & 0.012 &  & $-0.127$ & 0.035 \\
 \cufe  & 0.006 & 0.056 &  & $-0.307$ & 0.152 \\
 \znfe  & 0.002 & 0.042 &  & $-0.172$ & 0.061 \\
 \yba   & 0.002 & 0.031 &  & $-0.123$ & 0.066 \\
\noalign{\smallskip}
\hline
\end{tabular}
\end{table}

As seen from Table \ref{table:ratios}, the mean value of the 
deviations is close to zero for the group of high-$\alpha$ stars,
as one would expect from the definition of  $\Delta$, and the
scatter is not much higher than expected from the estimated
error of the abundance ratios. For the group of low-$\alpha$ 
stars the mean value of  $\Delta$ is always negative and
the scatter is significantly higher than the scatter for the
high-$\alpha$ group except in the case of Cr and Mn.

\begin{figure*}
\resizebox{\hsize}{!}{\includegraphics{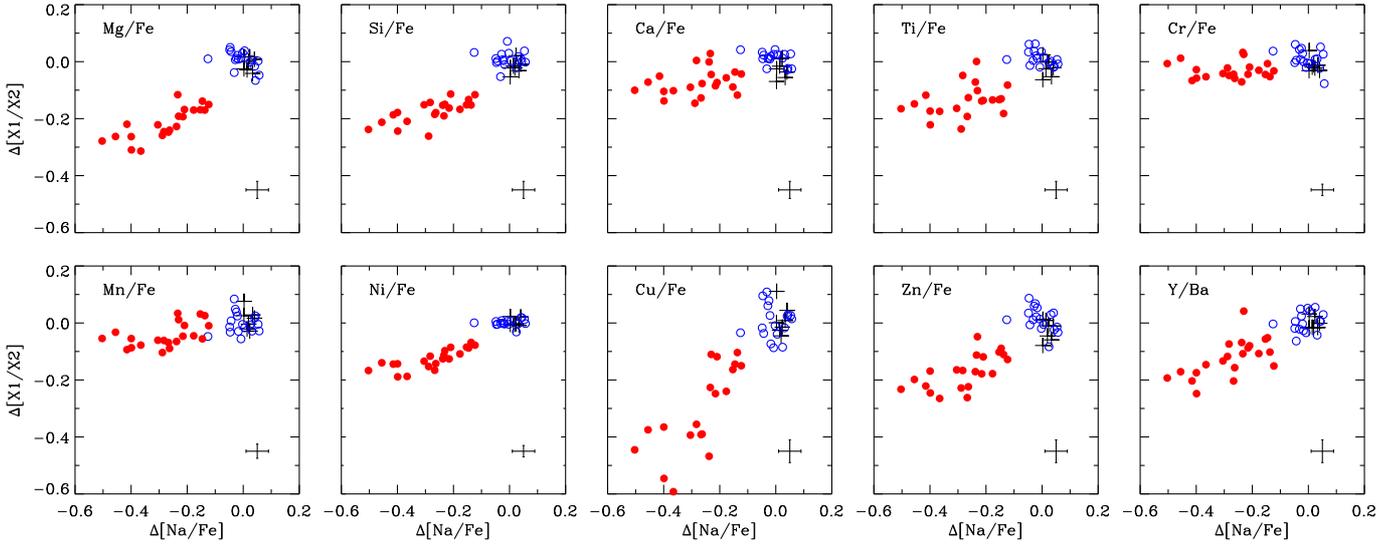}}
\caption{Correlation between differences in abundance ratios
relative to quadratic fits to the metallicity trends for
thick-disk and high-$\alpha$ halo stars (see Fig. \ref{fig:na.mg-fe}).
Only stars in the metallicity range $-1.1 \le \feh < -0.7$ are
plotted. The same symbols as in Fig. \ref{fig:mn.cu.zn-fe} are
employed.}
\label{fig:dX-dNa.limfe}
\end{figure*}

As shown in Fig. \ref{fig:dX-dNa.limfe}, there are remarkable
correlations between the deviations in the various abundance
ratios. Choosing $\Delta \nafe$ as the abscissa, one sees
very significant correlations for \mgfe , \sife , \tife ,
\nife , \cufe , \znfe , and \yba . The correlation 
is less clear in the case of \cafe , and the distributions of
$\Delta \crfe$ and $\Delta \mnfe$ are nearly flat.

\begin{figure}
\resizebox{\hsize}{!}{\includegraphics{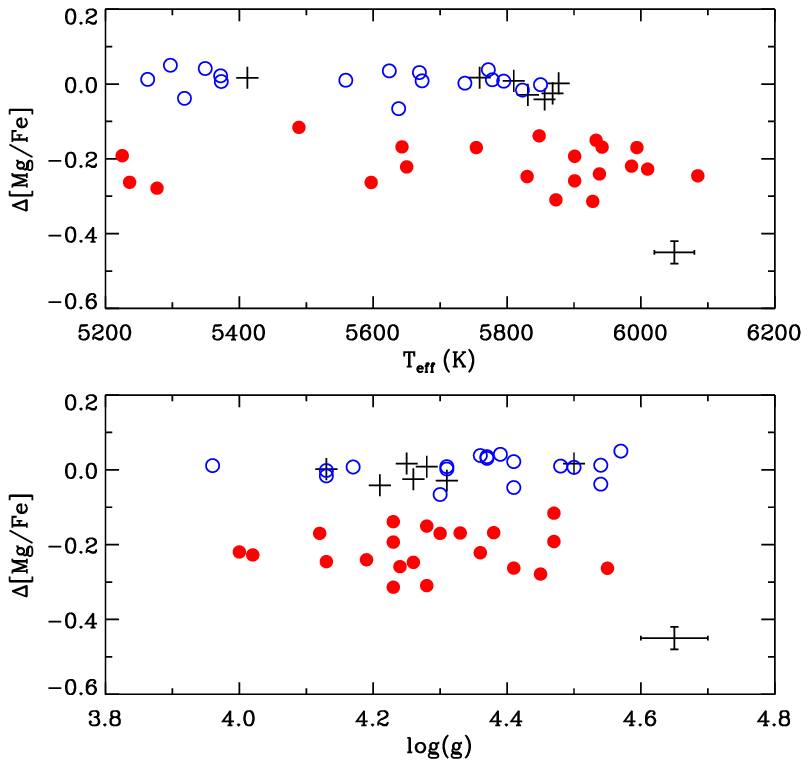}}
\caption{$\Delta \mgfe$ vs. \teff\ and \logg\ for stars in
the metallicity range $-1.1 \le \feh < -0.7$.
The same symbols as in Fig. \ref{fig:mn.cu.zn-fe} are
employed.}
\label{fig:dMg-Teff.logg}
\end{figure}

One may ask if the abundance differences between high- and
low-$\alpha$ stars, and the correlations seen in
Fig. \ref{fig:dX-dNa.limfe} could be spurious, due to errors
in \teff\ and/or \logg . This is, however, not the case.
As shown in Fig. \ref{fig:dMg-Teff.logg}, $\Delta \mgfe$
 has no significant dependence on \teff\ or \logg\, and the same
is the case for the other abundance ratios. Furthermore, 
the distribution of stars as a function of \teff\ and \logg\
is similar for the high- and low-$\alpha$ populations, except
for one interesting detail; the maximum \teff\ for
the low-$\alpha$ stars is approximately 150\,K higher than
the maximum for the high-$\alpha$ group. These maxima are
likely to correspond to the turnoff-points in the HR-diagram
for the two populations, and depend on the ages of the 
stellar populations as well as \feh\ and \alphafe . Using
the position of stars in the \logg\ -- \teff\ diagram
to derive individual ages, Schuster et al. (in preparation)
find that in fact stars in the low-$\alpha$ population
are on the average 2-3\,Gyr younger than stars belonging 
to the high-$\alpha$ population.

\subsection{Comparison with nucleosynthesis calculations}
\label{sect:nucleosynthesis}

\subsubsection{The $\alpha$-capture elements and sodium}
\label{sect:alpha-na}
Magnesium is a product of hydrostatic carbon and neon burning
in massive stars and is dispersed to the interstellar medium
by SNe\,II (e.g. Woosley \& Weaver \cite{woosley95}).
Very little Mg is produced by SNe\,Ia 
(Nomoto et al. \cite{nomoto97}). Our comparison
element, iron, partly comes from explosive oxygen and silicon
burning in SNe\,II, but is also produced by SNe\,Ia;
according to Tsujimoto et al. (\cite{tsujimoto95}) about 
60\% of iron in the Sun originates from these less massive
supernovae. The downward trend of \mgfe\ with increasing
metallicity for the low-$\alpha$
population may therefore be explained by delayed 
contribution of iron from SNe\,Ia, as first suggested by
Tinsley (\cite{tinsley79}) in a discussion of the trend of
\ofe\ vs. \feh\ for Galactic halo and disk stars.

The near-constancy of \mgfe\ for the
high-$\alpha$  population indicates that these
stars have been formed in regions with a relatively high SFR,
such that only SNe\,II have contributed to the chemical evolution
up to $\feh \sim -0.4$. The low-$\alpha$ stars, on the other hand,
originate from systems with a slower chemical evolution, where
SNe\,Ia have started to contribute iron at a metallicity 
$\feh \simlt -1.5$.  
Good candidates for such systems are
dwarf galaxies, for which the chemical evolution is thought
to proceed slowly because of a low SFR and/or loss of produced
elements in galactic winds (e.g. Lanfranchi \& Matteucci
\cite{lanfranchi03}; Fenner et al. \cite{fenner06}).

Sodium is produced by hydrostatic carbon burning in massive stars
with a yield that depends on the neutron excess
(e.g. Kobayashi et al. \cite{kobayashi06}).
Neutrons are released
by the $^{22}$Ne($\alpha$,n)$^{25}$Mg reaction, where $^{22}$Ne
comes from double $\alpha$-capture on $^{14}$N, which is enhanced in
the CNO cycle during hydrogen burning at the expense of carbon and
oxygen. As a result, the efficiency of
Na production increases with the  original C and O abundances of the star. 
This explains
the rising trend of \nafe\ vs. \feh\ for the high-$\alpha$ stars
in Fig. \ref{fig:na.mg-fe}. SNe\,Ia do not make Na in any significant amount  
(Nomoto et al. \cite{nomoto97}), so the underabundance of
Na relative to Fe in low-$\alpha$ stars, and the correlation
between $\Delta \nafe$ and $\Delta \mgfe$ (Fig. \ref{fig:dX-dNa.limfe})
is partly due to the excess of iron produced by SNe\,Ia. The amplitude of
the variations in \nafe\
is, however, significantly larger than in the case of \mgfe\ 
(Table \ref{table:ratios}). We suggest
that this is connected to the CO dependent yield of Na;
at a given \feh , the low-$\alpha$ massive stars synthesizing
Na contain less carbon and oxygen than massive stars belonging to the 
high-$\alpha$ population\footnote{We have not yet made a quantitative
analysis of the abundances of C and O, but a comparison of spectra for stars
with similar \teff , \logg , and \feh\ shows that low-$\alpha$ stars
have weaker \CI\ and \OI\ lines than high-$\alpha$ stars.}.

In addition to Na production in massive stars, sodium may also be
made in hydrogen burning shells of intermediate mass  AGB stars
in connection with the CNO and Ne-Na cycles (Ventura et al.
\cite{ventura01}). This process is often assumed to
be responsible for the high Na abundances and the Na--O anti-correlation
in globular cluster stars (Carretta et al. \cite{carretta09}).
If this process had contributed significantly
to the chemical evolution of the low-$\alpha$ population, one would not
expect such a clear correlation between \nafe\ and \mgfe\ as seen in
Fig. \ref{fig:dX-dNa.limfe}. As mentioned in NS10, there are, however, 
two  Na-rich stars among the  low-$\alpha$ population
(\object{G\,53-41} with \nafe = 0.23 and \object{G\,150-40} with
\nafe = 0.28) falling outside the boundaries of 
Figs. \ref{fig:na.mg-fe} and \ref{fig:dX-dNa.limfe}. These
stars could be halo field counterparts of the Na-enhanced 
globular cluster stars, formed in the vicinity of an intermediate
mass AGB star before its nucleosynthesis products were mixed into
the interstellar medium.

The remaining $\alpha$-capture elements, Si, Ca, and Ti are made
by oxygen and silicon burning in massive stars, but also have
a significant contribution from SNe\,Ia. According to
Tsujimoto et al. (\cite{tsujimoto95}, Table 3),
the relative contributions of SNe\,Ia to the solar abundances
are negligible for Mg, 17\% for Si, 25\% for Ca, and
57\% for Fe. This explains why the amplitudes of the variations
in \sife\ and \cafe\ are smaller than in the case of \mgfe\ 
(Table \ref{table:ratios}). It is, however, puzzling that the
variation in \cafe\ is so small compared to the variation in
\mgfe\ (only about one-third). Titanium was not included 
in the calculations of Tsujimoto et al. (SNe\,II models underproduce
Ti by a factor $\simgt \! 2$, e.g. Kobayashi et al. \cite{kobayashi06}),
but according to Table \ref{table:ratios} and Fig. \ref{fig:dX-dNa.limfe}, the
amplitude of the \tife\ variations is larger than that of \cafe . 

\subsubsection{Chromium and manganese}
\label{sect:cr-mn}
Chromium is produced by both Type II and Ia SNe. As seen
from Fig. 6 in NS10, [Cr/Fe] is close to zero and
there is no significant difference
in [Cr/Fe] between high- and low-$\alpha$ stars. Thus, chromium 
follows iron closely. This agrees with the
calculations of Kobayashi et al. \cite{kobayashi06}, who
include nucleosynthesis contributions from both normal SNe\,II
with explosion energy $\sim \! 10^{51}$\,erg,
and more energetic core collapse SNe, so-called hypernovae (HNe)
with 10 times higher explosion energy. For masses above 
20\,$M_{\odot}$, the fraction of 
of HNe is assumed to be 0.5. In addition, contribution from
SNe\,Ia is taken into account with yields from Nomoto et al.
(\cite{nomoto97}). The calculated \crfe\ is close to zero for 
metallicities ranging from $\feh \simeq -3$ to solar.

Like in the case of \crfe , there is no significant difference in 
\mnfe\ between high- and low-$\alpha$ stars, but  
\mnfe\ increases from about $-0.4$\,dex at
$\feh = -1.6$ to $-0.2$\,dex at $\feh = -0.4$. This trend continues 
for disk stars so that $\mnfe \simeq 0$ at solar metallicity
(e.g. Feltzing et al. \cite{feltzing07}).

Gratton (\cite{gratton89}) suggested that the rise in \mnfe\ is
caused by an overproduction of Mn relative to Fe in SNe\,Ia
as also predicted from the yields of Nomoto et al. (\cite{nomoto97}).
Accordingly, one would expect the low-$\alpha$ halo
stars to lie above the \mnfe\ trend defined by the high-$\alpha$ stars, but
this is not the case as seen from Fig. \ref{fig:mn.cu.zn-fe}. The 
explanation may be that the Mn yields of SNe\,Ia are metallicity dependent
as suggested from models by Badenes et al. (\cite{badenes08}). They find that
the increase in neutron excess with metallicity leads to a 
faster increase of Mn/Cr in Type Ia SNe than in core collapse SNe.
Recalling that the SNe\,Ia which enrich the low-$\alpha$ population are
relatively metal-poor, this may cancel the expected overabundance
of \mnfe\ for the low-$\alpha$ population. In extreme systems such as
the Sagittarius dwarf spheroidal (dSph) galaxy, the metallicity dependent Mn yield 
of SNe\,Ia may even lead to a decrease of \mnfe\ relative to
the trend for disk stars as first suggested by
McWilliam et al. (\cite{mcwilliam03}) and later shown in detail by
Cescutti et al. (\cite{cescutti08}) on the basis of chemical evolution
models.

Kobayashi \& Nomoto (\cite{kobayashi09}) have argued that if the
iron abundance of the progenitors of Type Ia SNe is 
lower than $\feh \sim -1$, then the wind of the white dwarf in a 
binary system is too weak for an explosion to occur. The absence
of SNe\,Ia below $\feh \sim -1$ could explain that the trend
of \mnfe\ is the same for high- and low-$\alpha$ stars, but then
one has to assume different initial mass functions (IMFs) for the
two populations in order to explain the differences in 
\alphafe . According to
Kobayashi et al. (\cite{kobayashi06}), an IMF biased towards
low-mass SNe\,II would lead to lower values of \alphafe , but because
the difference in \alphafe\ between the high- and low-$\alpha$
populations increases with \feh , one has to assume a very
complicated behavior of the IMF, i.e. a metallicity
dependent IMF for the low-$\alpha$ stars and a constant IMF
for the high-$\alpha$ stars. 

The rise in \mnfe\ with \feh\ for the high-$\alpha$ population
cannot be explained as caused by SNe\,Ia,
because the constancy of \mgfe\ excludes any significant
contribution from SNe\,Ia. Instead, the rise could be due to a
metallicity-dependent yield of Mn in SNe\,II, but the calculated yield
ratio Mn/Fe show only a relatively weak increase with metallicity
starting at $\feh \sim -1.2$ (Kobayashi et al. \cite{kobayashi06}, Fig. 5).
Perhaps our assumption of LTE is leading to a spurious increase in 
\mnfe\ with \feh . According to Bergemann \& Gehren (\cite{bergemann08}),
the strengths of \MnI\ lines are affected by departures from LTE;
the non-LTE corrections on the derived Mn abundances
are positive and increase with decreasing metallicity. When taken this
into account, the trend of \mnfe\ becomes more flat and approaches
$\mnfe \sim 0$ at all metallicities.

\subsubsection{Nickel and copper}
\label{sect:ni-cu}
It was long thought that nickel closely follows iron
(i.e. $\nife \simeq 0$) in both halo and disk stars.
Nissen \& Schuster (\cite{nissen97}) found, however, \nife\ to be
slightly negative in eight halo stars having unusually
low values of \alphafe\ and \nafe . Even more negative values
of \nife\ have been found in dSph galaxies (Venn et al. \cite{venn04}).
The underabundance of \nife\ for low-$\alpha$ stars
is confirmed in NS10, and as seen from Fig. \ref{fig:dX-dNa.limfe},
there is a very tight correlation between $\Delta \nife$ and
$\Delta \nafe$ for stars in the range $-1.1 \le \feh < -0.7$.

As discussed in Venn et al. (\cite{venn04}), the correlation between
Na and Ni, can be explained if the production of  $^{58}$Ni
(the most abundant Ni isotope) in SNe\,II depends on the neutron
excess in the same way as $^{23}$Na, i.e. the yields of 
both elements are metallicity dependent. Ni is, however, also produced by
SNe\,Ia. Current models of SNe\,Ia (e.g. Nomoto et al. \cite{nomoto97})
predict an overproduction of Ni relative to Fe, and it is therefore
puzzling that the low-$\alpha$ stars fall below the high-$\alpha$ 
group in the \nife\ - \feh\ diagram. Evidently, there is something
wrong with the yield calculations. As noted by  
Kobayashi et al. (\cite{kobayashi06}), the Ni yield from SNe\,Ia
depends strongly on the electron excess, $Y_e$, in the burning
region, which is sensitive to uncertain parameters such as the 
propagation speed of the burning front and the central density
of the white dwarf progenitor. Hence, there seems to be room for a downward
revision of Ni yields for SNe\,Ia.

The increase in \cufe\ with \feh\ (Fig. \ref{fig:mn.cu.zn-fe}) for
the high-alpha stars, can be explained if copper is mainly 
produced in massive stars by the weak $s$-process
(e.g. Bisterzo et al. \cite{bisterzo04}). The necessary neutrons
come from the $^{22}$Ne($\alpha$,n)$^{25}$Mg reaction,
which also controls the production of Na. Thus, the efficiency of
Cu production by the weak $s$-process also increases with increasing
original CNO abundance. This is confirmed by
detailed calculations of yields by Kobayashi et al. (\cite{kobayashi06}),
who find a steep increase in Cu/Fe as a function of heavy element abundance.

Current SNe\,Ia models predict negligible Cu yields
(e.g. Nomoto et al. \cite{nomoto97}).  The reduced
Cu/Fe ratio in low-$\alpha$ stars therefore arises because they
were formed from gas enriched with Fe from SNe\,Ia
and with Cu produced by CNO-poor massive stars. The same type of explanation
was suggested in Sect. \ref{sect:alpha-na} for the underabundance 
of \nafe\ in the low-$\alpha$ population.  As seen from
Fig. \ref{fig:dX-dNa.limfe}, there is indeed a nice correlation
between $\Delta \cufe$ and $\Delta \nafe$ with a slope that is
close to one.

\subsubsection{Zinc}
\label{sect:zinc}
The distribution of stars in the \znfe\ - \feh\ diagram 
(Fig. \ref{fig:mn.cu.zn-fe}) suggests that zinc behaves in much
the same way as an $\alpha$-element. \znfe\ is near-constant at
a level of +0.15\,dex for the thick-disk and high-$\alpha$ halo stars,
whereas the low-$\alpha$ stars show a declining trend in \znfe\
as a function of increasing \feh\ with an amplitude similar to
that of \sife\ (Table \ref{table:ratios}). 

The heavy
isotopes, $^{66-70}$Zn, are probably made by the weak $s$-process
in massive stars (Bisterzo et al. \cite{bisterzo04}). According to
Kobayashi et al. (\cite{kobayashi06}), the most abundant isotope,
$^{64}$Zn, is mainly formed by Si-burning in hypernovae. Type Ia SNe
produce relatively little Zn. Hence, the decreasing trend of
\znfe\ in the low-$\alpha$ population may be explained by the
production of iron in SNe\,Ia.

One may ask if the derived overabundance of Zn relative to Fe 
could be a spurious result because of the assumption of LTE.
According to the statistical equilibrium calculations for 
the energy level populations of the \ZnI\ atom
by Takeda et al. (\cite{takeda05}), the non-LTE corrections of
Zn abundances derived from the $\lambda \lambda 4722.1, 4810.5$
lines increase with decreasing
metallicity and reach about +0.1\,dex relative to the Sun
for a main-sequence star with $\teff \simeq 5800$\,K and 
$\feh \simeq -1$. This is of the same order of size as
the non-LTE correction for the Fe abundance derived from \FeI\
lines (as estimated from the difference in Fe abundances
derived from \FeII\ and \FeI\ lines; see Sect. \ref{sect:methods}).
There is some uncertainty in the statistical calculations due to a poor 
knowledge of the cross section for collisions between Zn and H atoms,
but altogether it is unlikely that the inclusion of non-LTE corrections would
lead to systematic changes in the derived \znfe\ values
by more than $\pm 0.1$\,dex,
and the difference in \znfe\ between the high- and low-$\alpha$ stars 
at a given metallicity would not be affected.  

When studying elemental abundances in the interstellar
gas of damped Lyman-alpha
systems (DLAs), it is often assumed that Zn can
be used as a proxy for Fe.  In support of this assumption, reference
is given to the study of Zn abundances in halo and disk stars 
by Sneden et al. (\cite{sneden91}). Zinc has the advantage of being a
volatile element like sulfur, and is therefore not depleted
onto interstellar grains. Thus, it has been suggested
that the S/Zn ratio can be used to date the star formation
process in DLAs in the same way as $\alpha$/Fe for stars. 
According to the present work Zn is, however, not
a good proxy for Fe. Supporting evidence comes from the work of
Bensby et al. (\cite{bensby05}), who found an offset in \znfe\
between thick- and thin-disk star amounting to about 0.15\,dex.
Altogether, this means that S/Zn is not a sensitive tracer of the star formation
history in galaxies. The same conclusion was reached by Nissen et al.
(\cite{nissen07}) from a direct study of [S/Zn] as a function \feh\
in Galactic stars.

\subsubsection{Yttrium and barium}
\label{sect:y-ba}

Yttrium and barium can be made by neutron capture onto
iron-peak nuclei by the weak $s$-process in massive stars
(like copper, see Sect. \ref{sect:ni-cu})
as well as the $r$-process, which is also thought to take
place in massive stars. From a Galactic evolution model,
Travaglio et al. (\cite{travaglio04}) predict that both
\yfe\ and \bafe\ are close to zero in halo and thick-disk
stars with $\feh > -2$. This is in rough agreement with 
the observed trend for high-$\alpha$ stars (Fig. \ref{fig:y.ba.bay-fe}),
although there is a puzzlingly large scatter
suggesting a poor degree of mixing of Y and Ba with Fe in the 
interstellar gas.

In addition, Y and Ba can be
made by the main $s$-process occurring
during shell He-burning in low-mass ($M \simlt 4 \, M_{\odot}$)
AGB stars (Busso et al. \cite{busso99}). 
Yttrium ($Z = 39$) belongs to the group of light $s$-process
($ls$) elements, whereas Ba ($Z = 56$) is one of the 
heavy $s$-process ($hs$) elements. The neutrons mainly
come from the $^{13}$C($\alpha, n$)$^{16}$O reaction,
where $^{13}$C is made by proton capture onto $^{12}$C nuclei
produced during He-burning. Hence, the neutron flux per seed
(iron) nuclei increases with decreasing metallicity. 
This means that low-metallicity AGB stars 
produce a higher $hs / ls$ ratio than 
high-metallicity AGB stars (Travaglio et al. \cite{travaglio04}).
The increasing trend of \bay\ 
for the low-$\alpha$ stars (Fig. \ref{fig:y.ba.bay-fe}) may therefore
be explained as due to delayed production of $s$-process elements
by metal-deficient, low-mass AGB stars. 

Given that the progenitors of 
low-mass AGB stars evolve on a timescale
ranging from $\sim 2 \times 10^8$ years to several billion years
(Schaller et al. \cite{schaller92}),
the \bay\ trends in Fig. \ref{fig:y.ba.bay-fe} support the idea that
the low-$\alpha$ stars have been formed in systems with a 
slower chemical evolution than the regions from which the high-$\alpha$
and thick-disk stars originated.  Detailed models for the
chemical evolution of dSph galaxies including mass-loss in
SNe winds (Fenner et al. \cite{fenner06}) confirm that an
increase in Ba/Y for the metallicity range $-2 < \feh < -1$ 
is obtained, when the star formation history extends over 
more than one Gyr.

\subsection{Comparison with dSph and Irr galaxies}
\label{sect:dwarfs}
As discussed in NS10, the kinematics of the low-$\alpha$
stars suggest that they have been accreted from Milky-Way
satellite galaxies. It is therefore interesting to compare
abundance ratios measured in such present-day galaxies
with the corresponding ratios for the low-$\alpha$ stars.

As reviewed by Tolstoy et al. (\cite{tolstoy09}), the knowledge
about elemental abundances in dwarf galaxies has been
greatly improved over the last few years thanks to the advent
of multi-object, high-resolution spectrographs at large telescopes
allowing the determination of abundances for
samples of red giant branch (RGB) stars.
As seen from Fig. 11 in Tolstoy et al. (\cite{tolstoy09}),
dSph stars with low metallicities, $\feh \simlt -2$, have 
about the same $\alpha$/Fe ratio as stars in the Galactic 
halo ($\alphafe \sim +0.3$), which points to a common
IMF. At some metallicity (the so-called
`knee' position, $\feh_{\rm knee}$), \alphafe\ starts to decrease 
as a function of increasing metallicity.
This is similar to 
the trend for the low-$\alpha$ field stars except that
the amplitude of the  \alphafe\ variations tends to be larger
for dSph galaxies than in the case of the low-$\alpha$
population in the solar neighborhood.

There is some evidence that $\feh_{\rm knee}$
differs among the dwarf spheroidals 
and is correlated with the maximum \feh\ reached, 
i.e. that metallicity at which the gas is blown out of the
system by SNe.  
The low-luminosity dSph systems, Carina, Draco, and Ursa Minor,
have $\feh_{\rm knee} \sim -2.4$ to $-2.0$, and
$\feh_{\rm max} \sim -1.5$ (Koch et al. \cite{koch08};
Cohen \& Huang \cite{cohen09}, \cite{cohen10}).  The more
luminous Sculptor dSph has $\feh_{\rm knee} \sim -1.8$,
and $\feh_{\rm max} \sim -1.0$ (Tolstoy et al. \cite{tolstoy09}).
In the case of the still more luminous Fornax system, the
`knee' position is not well defined, but $\feh_{\rm max} \sim -0.6$
(Letarte et al. \cite{letarte10}). Finally, the Sagittarius
dSph has $\feh_{\rm knee} \sim -1.3$, and $\feh_{\rm max} \sim 0.0$
(Sbordone et al. \cite{sbordone07}; Carretta et al. \cite{carretta10}).
This maximum metallicity occurs in the core of Sagittarius; 
the metallicity of stars in the trailing and leading arms 
(i.e. stars presently being accreted by the Galaxy) tend to be lower
(Chou et al. \cite{chou07}; Keller et al. \cite{keller10})

The `knee' position for the low-$\alpha$ population is somewhat
uncertain because no stars with $\feh < -1.6$ were included.
According to Fig. 1 in NS10, it seems to lie around or slightly below
a  metallicity of $-1.5$, i.e. $\feh_{\rm knee} \simlt -1.5$.
The maximum metallicity for the low-$\alpha$ stars, on the other hand, is
well-defined at $\feh_{\rm max} \simeq -0.75$. 
Hence, there is no exact correspondence with any of the
existing dSph systems, but this does not exclude that
the low-$\alpha$ stars originate from 
now dissolved dSph systems. 
The reason  for the scatter of \alphafe\ at a given metallicity
in the low-$\alpha$ population could be that they were accreted from
an ensemble of dwarf galaxies with somewhat different star formation
efficiencies.

\begin{figure}
\resizebox{\hsize}{!}{\includegraphics{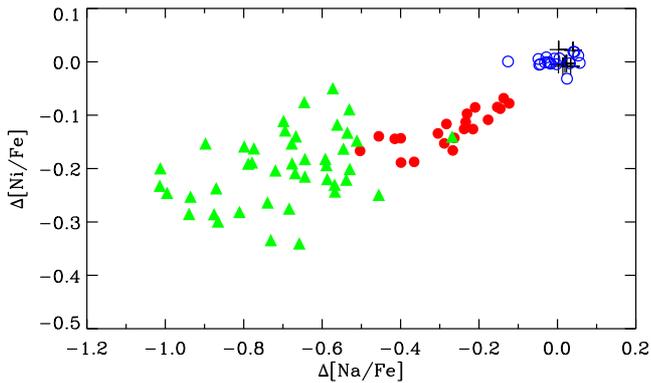}}
\caption{$\Delta \nife$ vs. $\Delta \nafe$ for stars in the
metallicity range $-1.1 \le \feh < -0.7$. Stars in this paper are
shown with the same symbols as in Fig. \ref{fig:mn.cu.zn-fe}.
Green triangles show
values for RGB stars in the Fornax dSph galaxy (Letarte et al.  
\cite{letarte10}).}
\label{fig:dNi-dNa.letarte}
\end{figure}

For a more detailed comparison of abundance ratios in dSph galaxies
and the low-$\alpha$ field stars, we may use Fornax
as a well studied example (Letarte et al. \cite{letarte10}).
\mgfe\ and \sife\ overlap with the corresponding ratios
for the low-$\alpha$ stars,
but \cafe\ and \tife\ are lower in the Fornax stars.
According to Letarte et al. (\cite{letarte10}), the derived
Ca and Ti abundances may, however, be affected by non-LTE effects.
Spurious offsets may also arise when comparing abundances for
two samples of stars with different sets of atmospheric parameters,
i.e. red giants in dSph galaxies and dwarf stars in the solar
neighborhood.

As seen from
Fig. \ref{fig:dNi-dNa.letarte}, \nafe\ and \nife\ in Fornax 
extend the correlation defined by the solar neighborhood stars to
lower values of \nafe , albeit with a larger scatter because of
the lower precision of the Fornax abundances. 
\bay\ in Fornax rises sharply with increasing \feh\ and reaches $\bay \sim 0.8$
at $\feh = -0.8$, i.e. much higher than \bay\ for the low-$\alpha$
stars. Thus, Fornax has more extreme
abundance anomalies than the low-$\alpha$ population. This may be 
related to differences in the SFR; the majority of stars in
Fornax have ages between 2 and 6 Gyr (Letarte et al. \cite{letarte10}),
whereas the low-$\alpha$ stars have ages around 10 Gyr 
(Schuster et al., in preparation).

Mn, Cu, and Zn were not included in the Fornax study by 
Letarte et al. (\cite{letarte10}). Sagittarius is the only dSph
for which these elements have been studied in some detail
and the data refer mainly to the metallicity range $-0.5 < \feh < 0.0$.
\mnfe\ falls below the trend for Galactic disk stars
(McWilliam et al. \cite{mcwilliam03}), and both \cufe\ and 
\znfe\ are even lower than in the
low-$\alpha$ population (McWilliam \& Smecker-Hane \cite{mcwilliam05};
Sbordone et al. \cite{sbordone07}). Again, one may speculate that these
extreme abundance anomalies are connected to a very prolonged
star formation history of Sagittarius.

It is also interesting to compare the abundance ratios in low-$\alpha$
stars with ratios obtained for irregular galaxies.
Kaufer et al. (\cite{kaufer04}) analyzed three A-type supergiants in
the dwarf Irr galaxy Sextans A and found
$\alphafe \sim -0.1$ at $\feh = -1.0$. The same low value
of \alphafe\ was obtained for three M-type supergiants
with $\feh \simeq -0.7$ in the dIrr galaxy IC\,1613 by
Tautvai$\breve{\rm s}$iene et al. (\cite{tautvaisiene07}).
Hence, \alphafe\ in these dIrr galaxies is  about 0.2 dex lower than the 
ratio for low-$\alpha$ halo stars. 

Finally, there are some similarities between
abundance ratios in low-$\alpha$ halo stars and 
ratios obtained by Pomp\'{e}ia et al. (\cite{pompeia08})
for RGB stars belonging to the inner disk of the
Large Magellanic Cloud (LMC). 
The trends of \mgfe , \sife , and \tife\ as a function of \feh\ 
are about the same except that metallicities as high as 
$\feh \sim -0.4$ are reached in LMC. 
\nafe\ and \nife\ are negative in the LMC stars,
similar to the low-$\alpha$ stars, but
the correlation between these ratios is not so clear as in 
Fig. \ref{fig:dNi-dNa.letarte} (see Fig. 14 in 
Pomp\'{e}ia et al. \cite{pompeia08}). Furthermore, underabundances
of \cufe\ are observed both in the LMC and in the low-$\alpha$
population. However, there also are differences;
the LMC stars have lower \cafe , and \bay\ shows a much steeper
increase with \feh\ than found for the low-$\alpha$ stars.

\subsection{Comparison with $\omega$\,Cen}
\label{sect:omegacen}
As mentioned in Sect. \ref{sect:introduction}, the majority of 
low-$\alpha$ stars move on retrograde orbits close to the Galactic plane
with a wide distribution of the $U$-velocity component.
This resembles the
orbital characteristics of stars captured from a
satellite galaxy that is dragged into the Galactic plane
by dynamical friction (Meza et al. \cite{meza05}).
Given that the $\omega$\,Cen globular cluster is probably
the nucleus of a dwarf galaxy
(e.g. Bekki \& Freeman \cite{bekki03}), and that
many of the low-$\alpha$ stars have a $V$-velocity component 
similar to that of $\omega$\,Cen, 
it is possible that a significant fraction of the low-$\alpha$
halo stars in the solar neighborhood have been formed in
the $\omega$\,Cen progenitor galaxy. Hence, it is of
interest to compare abundance ratios in the low-$\alpha$
stars with the corresponding ratios in $\omega$\,Cen stars.

$\omega$\,Cen is unique among globular clusters by having a
wide metallicity distribution.  
In a comprehensive, high-resolution
study of 855 red giants in $\omega$\,Cen, i.e. essentially
all stars brighter than $V = 13.5$, Johnson \& Pilachowski
(\cite{johnson10}) identified four peaks in the metallicity 
distribution at $\feh \simeq -1.75$, $-1.50$, $-1.10$, and $-0.75$.
A Gaussian decomposition of the distribution (see Fig. 8 in
Johnson \& Pilachowski \cite{johnson10}) indicates a
corresponding frequency distribution of 61\,\%, 27\,\%, 10\,\%, and 2\,\%, 
for the stars in $\omega$\,Cen. The upper limit of 
\feh\ in  $\omega$\,Cen corresponds quite well to that of the 
low-$\alpha$ stars, but there are too few metal-rich stars 
in $\omega$\,Cen to match the \feh\ distribution for the
low-$\alpha$ stars stars. 

Regarding the trend of \alphafe\ vs. \feh\ in $\omega$\,Cen,
there are conflicting results. From high-resolution spectroscopy 
of six red giants, Pancino et al. (\cite{pancino02}) found 
evidence of a decline of \sife\ and \cafe\ starting around
$\feh \sim -1$. A similar decline in \alphafe\ 
($\alpha \equiv$ O, Mg, Si, and Ca)  was found by Origlia et al. 
(\cite{origlia03}) on the basis of medium-resolution IR spectra
of 21 giants in $\omega$\,Cen. The extensive high-resolution
study of Johnson \& Pilachowski (\cite{johnson10}), which
includes more than 70 stars in the metallicity range
$-1.2 < \feh < -0.7$, indicates, however, that \sife , \cafe , 
and \tife\ are essentially constant at a level of +0.3 to +0.4 dex
suggesting that SNe\,Ia have not contributed to the
chemical enrichment in $\omega$\,Cen. Even if one
believes in the results of Pancino et al. (\cite{pancino02}) 
and  Origlia et al. (\cite{origlia03}), it is clear that
the contribution of  SNe\,Ia started at a higher metallicity
($\feh_{\rm knee} \simeq -1$) than in the case of the present low-$\alpha$
population ($\feh_{\rm knee} \simlt -1.5$).
 
\nafe\ is very different in low-$\alpha$ and $\omega$\,Cen
stars. Instead of being negative, \nafe\ is positive in 
$\omega$\,Cen stars with $\feh > -1.4$, and there is a clear
Na--O anti-correlation (Johnson \& Pilachowski \cite{johnson10})
as found in all well-studied globular clusters 
(Carretta et al. \cite{carretta09}). This anti-correlation may be ascribed
to the CNO and Ne-Na cycles in connection with hydrogen
burning in intermediate-mass AGB stars (Ventura et al. \cite{ventura01}).
According to hydrodynamical
simulations by D'Ercole et al. (\cite{dercole08}), the gas ejected
from such AGB stars collects in the cluster core,
which could explain the difference in \nafe\ between stars remaining in
$\omega$\,Cen itself and those originating from the progenitor galaxy.

\begin{figure}
\resizebox{\hsize}{!}{\includegraphics{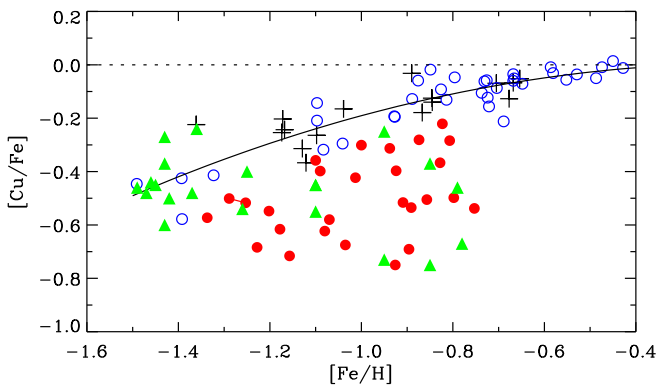}}
\caption{\cufe\ vs. \feh . Stars in this paper are
shown with the same symbols as in Fig. \ref{fig:mn.cu.zn-fe}.
The solid line is a quadratic fit to the \cufe\ - \feh\ trend
for thick-disk and high-$\alpha$ halo stars. Green triangles show
values for RGB stars in the $\omega$\,Cen globular cluster
(Cunha et al. \cite{cunha02}).}
\label{fig:cu-fe.cunha}
\end{figure}

Turning to the iron-peak elements, there are some similarities
between $\omega$\,Cen  and the low-$\alpha$ halo population.
As noted by Johnson \& Pilachowski (\cite{johnson10}),
\nife\ in $\omega$\,Cen decreases slightly at 
$\feh > -1.3$. From data in their
Table 5, an average $<\! \nife \!> \, = -0.11$ is obtained
for 42 stars with $-1.1 \le  \feh < -0.7$, which compares well
with the corresponding value $<\! \nife \!> \, = -0.13$ for 
21 low-$\alpha$ stars. As shown in Fig. \ref{fig:cu-fe.cunha},
there is also a striking agreement between
\cufe\ in low-$\alpha$ stars and the ratios determined 
for red giants in $\omega$\,Cen by Cunha et al. (\cite{cunha02}). 
Recently, Cunha et al. (\cite{cunha10}) have also determined
Mn abundances in ten RGB stars in $\omega$\,Cen covering the metallicity range
$-1.8 < \feh < -0.9$. Eight stars have \mnfe\ values overlapping the
trend for the solar neighborhood stars, but the two most
metal-rich stars with $\feh \simeq -1.2$ and $-0.9$ have
$\mnfe \simeq -0.9$, i.e. far below the values for the 
high- and low-$\alpha$ halo stars. More data on Mn abundances
in $\omega$\,Cen are needed to confirm, analyze, and understand this puzzling
deviation.

Abundances of neutron capture elements in $\omega$\,Cen
were determined for ten RGB stars having $-1.8 < \feh < -0.8$
by Smith et al. (\cite{smith00}).  In this metallicity range,
the abundances of the heavy-$s$ elements, Ba and La,
increase by more than 1\,dex relative to Fe.
\yfe\ shows a much smaller increase, i.e. less than 0.3\,dex. 
Hence, the $\omega$\,Cen stars show a much
more dramatic increase in \bay\ with \feh\ than the low-$\alpha$
stars. The reason for this difference could be
that winds from low-mass AGB stars producing the
$s$-process elements are more efficiently retained in the 
cluster than the faster moving Type II ejecta 
(Smith et al. \cite{smith00}). These authors also suggest
that all elements produced by Type Ia SNe escape from
the $\omega$\,Cen cluster, which would explain the `high'
trend of \alphafe\ found by Johnson \& Pilachowski
(\cite{johnson10}).
  
As seen from this discussion, only \nife\ and \cufe\ 
agree between the stars in the $\omega$\,Cen cluster
and those in the low-$\alpha$
population. There are large differences in the
distributions of  \alphafe , \nafe , and \bay .
Hence, if some of the low-$\alpha$ stars were 
accreted from $\omega$\,Cen,
the chemical evolution pattern must have been different
in the inner and outer parts of the progenitor galaxy.
Evidence of such a scenario comes from the recent work by 
Chou et al. (\cite{chou10}) on abundances in the Sagittarius
dSph galaxy; stars in the leading north arm tend to have 
lower values of [La/Y] than stars in the Sgr core for 
the metallicity range $-1 < \feh < 0$ (see Fig. 6
in the Chou et al. paper).

\section{Conclusions}
\label{sect:conclusions}

In this paper very precise abundances of Mn, Cu, Zn, Y, and Ba have
been determined relative to Fe for stars in the solar neighborhood
with halo kinematics. The two populations, high-
and low-$\alpha$ stars, that were found to have different
ratios of \alphafe , \nafe , and \nife\ in our first paper
(NS10) are also separated in \cufe , \znfe , and \bay .
There is, however, no significant 
differences in \mnfe . 

Trends and correlations between the abundance ratios may
to a large extent be explained from existing nucleosynthesis
calculations if the high-$\alpha$ stars have been 
formed in regions with such a high SFR that only
massive stars and SNe\,II have contributed to the
chemical evolution up to $\feh \simeq -0.4$.
The low-$\alpha$ stars, on the other hand, 
may originate from systems with a slower chemical evolution,
characterized by delayed enrichment from relatively metal-poor
SNe\,Ia and low-mass AGB stars in addition to the contributions
from massive stars and Type II SNe. It is, however, difficult to
explain that there is no significant difference in \mnfe\
between the high- and low-$\alpha$ populations.  Furthermore, the data
for \nife\ call for a revision of the Ni yields from SNe\,Ia.  

RGB stars in present-day dSph satellite galaxies show abundance
trends of  \alphafe , \nafe , \nife , \cufe , and \bay\ that
have some similarities with the trends for the low-$\alpha$ population,
but there is no exact matching. In general, existing 
dSph galaxies seem to be characterized by an even slower chemical
evolution than the systems in which the low-$\alpha$ stars were formed.

As discussed in NS10,
the kinematics of the low-$\alpha$ stars suggest  that some of
them were accreted from the progenitor galaxy of the $\omega$\,Cen
globular cluster. We have looked for supporting
evidence from chemical abundance ratios, but have found more
differences between the $\omega$\,Cen cluster and the
low-$\alpha$ stars than similarities. The distributions
of \nife\ and \cufe\ for RGB stars in $\omega$\,Cen
overlap well with those of the low-$\alpha$ population,
but \nafe\ and \bay\ are very different in $\omega$\,Cen and
low-$\alpha$ stars. This difference could be explained if
the products of AGB stars are selectively retained within the
$\omega$\,Cen cluster. The trend of \alphafe\ is also 
different for $\omega$\,Cen and the low-$\alpha$
population, so one has to invoke differential loss of SNe\,II
and SNe\,Ia winds from $\omega$\,Cen if some of the 
low-$\alpha$ stars did originate from the progenitor galaxy.

Zolotov et al. (\cite{zolotov09}, \cite{zolotov10}) have recently
used N-body and smooth particle hydrodynamic simulations to
investigate the kinematics and \alphafe\ trends of stellar halos
of large galaxies similar to the Milky Way. They
find that the inner halos ($R < 20$\,kpc) contain both accreted
and in situ formed stars. These two populations are
separated in \alphafe\ at the high end of the metallicity
distribution function if the halo was formed in connection with
a few major 
mergers ($M_{\rm satellite}/ M_{\rm primary} \, > 0.1$) at early times,
i.e. more than 8 -- 9 Gyr ago. The in situ halo stars formed 
in the innermost $\sim 4$\,kpc of the galaxy in a deep potential
well causing the SFR to be so high that only core collapse SNe
contributed to the chemical enrichment. Later these stars were
`heated' to halo kinematics by mergers. The accreted stars,
on the other hand, formed in satellite galaxies with shallower
potential wells and hence lower SFR allowing SNe\,Ia to 
contribute to the enrichment at the high end of the
metallicity distribution.

The dual distribution of \alphafe\ obtained in the
simulations of Zolotov et al. (\cite{zolotov10})
corresponds qualitatively to the abundance trends determined in
NS10 and in this paper. Thus,
the high-$\alpha$ population may consist of stars formed
in the innermost part of the Milky Way and displaced to the halo by mergers,
whereas the low-$\alpha$ stars may have been accreted at early 
times from a few, relatively massive satellite galaxies. 
The dispersion in abundance ratios at a given \feh\ 
suggests that these satellites followed somewhat different
chemical evolution patterns.
Present-day dSph galaxies are less massive, which may explain
that they have more extreme abundance anomalies than the 
low-$\alpha$ stars. Clearly, this scenario should be tested
by determining \alphafe\ and other abundance ratios for a large 
sample of halo stars situated in other regions than the 
solar neighborhood.

\begin{acknowledgements}
This publication made use of the SIMBAD database operated
at CDS, Strasbourg, France, and NASA's Astrophysics Data System.
The staff at the Nordic Optical Telescope is thanked for competent 
assistance in obtaining spectra for this project.
\end{acknowledgements}

\Online

\end{document}